\documentclass{article}
\usepackage{times}
\usepackage{geometry}
\usepackage[utf8]{inputenc}
\usepackage{enumitem,amssymb}
\usepackage{ragged2e}
\usepackage{graphicx}
\usepackage{comment}
\usepackage{multicol}
\usepackage[usenames]{xcolor} 
\usepackage{subfigure}
\usepackage{float}

\usepackage{amsmath}

\DeclareUnicodeCharacter{2009}{\,} 
\DeclareUnicodeCharacter{2212}{\-}

\definecolor{xlinkcolor}{cmyk}{1,1,0,0}
\usepackage{url}
\usepackage[
 colorlinks=true,    
 linkcolor=xlinkcolor,     
 citecolor=xlinkcolor,     
 filecolor=xlinkcolor,  
 urlcolor=xlinkcolor,      
 final=true
]{hyperref}

\usepackage{enumitem}
\setenumerate{itemsep=0mm}

\usepackage{authblk}
\title{\Large Low Background kTon-Scale Liquid Argon Time Projection Chambers}
\author[1]{A. Avasthi}
\author[2]{T. Bezerra}
\author[2]{A. Borkum}
\author[3]{E. Church}
\author[4]{J. Genovesi}
\author[4]{J. Haiston}
\author[3]{C. M. Jackson}
\author[5]{I. Lazanu}
\author[1]{B. Monreal}
\author[3]{S. Munson}
\author[6]{C. Ortiz}
\author[5]{M. Parvu}
\author[2]{S. J. M. Peeters}
\author[6]{D. Pershey}
\author[3]{S. S. Poudel}
\author[4]{J. Reichenbacher}
\author[3]{R. Saldanha}
\author[6]{K. Scholberg}
\author[4]{G. Sinev}
\author[7]{J. Zennamo}
\author[3]{H. O. Back}
\author[8]{J. F. Beacom}
\author[9]{F. Capozzi}
\author[10]{C. Cuesta}
\author[11]{Z. Djurcic}
\author[12]{A. C. Ezeribe}
\author[10]{I. Gil-Botella}
\author[7]{S. W. Li}
\author[13]{M. Mooney}
\author[9]{M. Sorel}
\author[14]{S. Westerdale}
\affil[1]{Case Western Reserve University, Cleveland, Ohio 44106, USA}
\affil[2]{University  of  Sussex,  Brighton,  BN1  9RH,  United  Kingdom}
\affil[3]{Pacific  Northwest  National  Laboratory,  Richland,  WA  99352,  USA}
\affil[4]{South  Dakota  School  of  Mines  and  Technology,  Rapid  City,  SD  57701,  USA}
\affil[5]{University of Bucharest, Bucharest, Romania}
\affil[6]{Duke  University,  Durham,  NC  27708,  USA}
\affil[7]{Fermi National Accelerator Laboratory, Batavia, IL 60510, USA}
\affil[8]{Ohio State University, Columbus, OH 43210, USA}
\affil[9]{Instituto de Física Corpuscular, 46980 Paterna, Valencia, Spain}
\affil[10]{CIEMAT, E-28040 Madrid, Spain}
\affil[11]{Argonne  National  Laboratory,  Argonne,  IL  60439,  USA}
\affil[12]{University of Sheffield, Sheffield, S3 7RH, United Kingdom}
\affil[13]{Colorado State University, Fort Collins, CO 80523, USA}
\affil[14]{Princeton University, Princeton, NJ 08544, USA}
\date{}                     
\begin{document}
\begin{raggedright} 
Snowmass2021 - White Paper\vspace{-2em} 
\end{raggedright}


\normalsize

{\let\newpage\relax\maketitle}
\vspace{-6mm}
\noindent {\large \bf NF Topical Groups:}

\noindent $\blacksquare$ (NF1) Neutrino oscillations \\
\noindent $\blacksquare$ (NF3) Beyond the Standard Model  \\
\noindent $\blacksquare$ (NF4) Neutrinos from natural sources\\
\noindent $\blacksquare$ (NF5) Neutrino properties \\
\noindent $\blacksquare$ (NF6) Neutrino cross sections \\
\noindent $\blacksquare$ (TF11) Theory of neutrino physics \\
\noindent $\blacksquare$ (NF9) Artificial neutrino sources \\
\noindent $\blacksquare$ (NF10) Neutrino detectors  \\
\noindent {\large \bf Other Topical Groups:} \\
\noindent $\blacksquare$ (CF1) Dark Matter: Particle-like \\
\noindent $\blacksquare$ (IF8) Noble Elements \\
\noindent $\blacksquare$ (UF01) Underground Facilities for Neutrinos \\
\noindent $\blacksquare$ (UF02) Underground Facilities for Cosmic Frontier \\
\noindent $\blacksquare$ (UF03) Underground Detectors \\

\noindent {\large \bf Contact Information:} \\
Eric Church (Pacific Northwest National Laboratory) [eric.church@pnnl.gov] \\
Christopher Jackson (Pacific Northwest National Laboratory) [christopher.jackson@pnnl.gov] \\
Juergen Reichenbacher (South Dakota School of Mines and Technology) [Juergen.Reichenbacher@sdsmt.edu] \\


\noindent {\large \bf Abstract:}
We find that it is possible to increase sensitivity to low energy physics in a third or fourth DUNE-like module with careful controls over radiopurity and some modifications to a detector similar to the DUNE Far Detector design. In particular, sensitivity to supernova and solar neutrinos can be enhanced with improved MeV-scale reach. A neutrinoless double beta decay search with $^{136}$Xe loading appears feasible. Furthermore, sensitivity to Weakly-Interacting Massive Particle (WIMP) Dark Matter (DM) becomes competitive with the planned world program in such a detector, offering a unique seasonal variation detection that is characteristic for the nature of WIMPs.

\clearpage

\section*{Executive Summary}
This white paper introduces the concept of a low background kTon-scale liquid argon time projection chamber that could be a possible design for the 3rd or 4th module of the Deep Underground Neutrino Experiment (DUNE). Such a module would allow the physics scope of that experiment to be increased with the addition of a number of low-energy physics topics, without disrupting the main oscillation physics program.

The detector design will take as a starting point the standard DUNE vertical drift single phase TPC design. It will be modified with the addition of an optically isolated inner volume, where fiducialization allows significantly lower background levels, and photosensor coverage can be increased to improve energy resolution. Radioactive backgrounds will be controlled by:
\begin{itemize}
    \item External neutrons from the cavern. A 40 cm layer of water shield will reduce external neutrons by a factor $10^{3}$.
    \item Cryostat and detector background. An intensive materials assay campaign will allow selection of construction materials with radioactive backgrounds reduced by a factor $10^{3}$. Such material purity has been exceeded by current dark matter experiments. Additional internal shielding can also be added, including neutron absorbers such as boron, lithium or gadolinium loaded cryostat layers.
    \item Radon in the liquid argon. Active purification of the inner volume and an emanation measurement campaign will reduce radon levels by a factor $10^{3}$. This factor has been exceeded by current dark matter experiments.
    \item Internal argon-42, argon-39, krypton-85. Underground argon from a viable commercial source will be used to reduce background isotopes found in conventional atmospheric argon sources.
\end{itemize}

The light collection will be enhanced in the inner volume with at least $\sim 10 \%$ coverage of SiPM tiles on walls and cathode, additional reflectors on the inner walls and anode collection planes, and additional argon purity requirements. This will allow energy resolution of $\sim 2\%$ at 1 MeV when combined with the TPC charge signal, and allow a pulse shape discrimination measurement for background reduction.

We show a number of important physics topics that can be explored with this detector:
\begin{itemize}
    \item Supernova neutrino physics. Lower neutron levels allow the lowering of energy threshold for the search to 600 keV, permitting access to interesting low energy and late or early time information. The reduction in backgrounds allows improvements to the trigger sensitivity, and the detector will be fully sensitive to supernova from the Magellanic cloud. In addition the increased light collection allows the supernova neutrino CE$\nu$NS interactions within the detector to be measured.
    \item Solar neutrino physics. Lowering the threshold and improving the energy resolution will increase sensitivity to $\Delta m^2_{21}$ allowing a precision measurement. This would also improve non-standard interaction constraints or explain the solar anomaly between reactor and solar measurements of $\Delta m_{21}^2$. A precision measurement of the CNO flux is also possible.
    \item Neutrinoless double beta decay. Loading the detector with a few-percent of xenon-136 allows a sensitive search beyond the coming ton-scale experiments
    \item Dark Matter (DM). The increased light coverage allows pulse shape discrimination to remove electron recoil backgrounds (primarily argon-39) to a Weakly Interacting Massive Particle (WIMP) dark matter search though nuclear recoils of at least 100 keV and perhaps just 50 keV. Such a search will be competitive with coming dark matter searches on a reduced timescale. Due to the unrivaled large mass of 3 ktons and a potentially very long DUNE operation of one decade (or even several), this concept can offer a unique seasonal variation detection at sufficient statistical significance for providing a smoking gun signature for the nature of WIMPs. This would be particularly of interest in case upcoming generation-2 DM experiments like LZ, XENONnT and/or DarkSide-20k have evidence for WIMPs near their sensitivity. It would be nearly impossible for the planned generation-3 DM experiments to make such a smoking gun detection proving the WIMP nature of DM. 
\end{itemize}

This concept takes full advantage of the planned DUNE far-detector facilities to create a true multipurpose detector to address the important questions in high energy physics, astronomy and nuclear physics.
\clearpage

\section{Introduction}

In this white paper we introduce and discuss a dedicated low background  module that would enhance the physics program of next-generation experiments such as the planned Deep Underground Neutrino Experiment (DUNE). Such a low background DUNE-like module could be installed as either module 3 or module 4, the so-called ``Module of Opportunity" in DUNE. Such a module would increase the physics reach of supernova and solar neutrino physics, and could potentially host a next generation neutrinoless double beta decay ($0\nu\beta\beta$) or Weakly Interacting Massive Particle (WIMP) dark matter search.

The physics reach would be enhanced by lowering the nominal energy threshold of a DUNE-like experiment from 5-10 MeV to three targets of increasing challenge:
\begin{itemize}
    \item $\sim$ 3.5 MeV energy threshold. This threshold is set by the $^{42}$K daughter of $^{42}$Ar in the detector target. By reducing neutron captures, alpha-emitting radon daughters and pileup events above this threshold, supernova burst neutrino sensitivity in distance, energy and time could be increased. Sensitivity to solar neutrinos would also be enhanced, allowing explorations of interesting solar-reactor tensions and Non-Standard Interactions.
    \item $\sim$ 0.5 MeV energy threshold. This threshold is set by the $^{39}$Ar in the detector target. With reduction of electron and photon backgrounds in the target, particularly if the $^{42}$Ar to $^{42}$K decay is reduced through use of underground argon (UAr), sensitivity to low energy solar neutrinos from the CNO process would allow a precision measurement to be made. With loading of $^{136}$Xe, a sensitive $0\nu\beta\beta$ search could also occur.
    \item $<100$ keV energy threshold. This threshold is achieved by enhancing the light collection within the detector and by lowering the $^{39}$Ar background by deploying UAr. With rejection of electron recoil backgrounds (using timing based pulse shape discrimination), a sensitive WIMP dark matter search could take place, and interesting phenomena such as a supernova coherent elastic neutrino-nucleus scattering signal (a CE$\nu$NS glow) could be studied.  
\end{itemize}
These low background targets are achievable due to the size of the module, allowing significant fiducialization and hence less stringent radioactive background requirements than current world-leading dark matter searches. The light enhancements are achievable with current production techniques. 

In this paper in Section~\ref{sec:DD} we outline the design of the module and discuss potential paths to achieve the detector requirements. In Section~\ref{sec:phys} we show our initial studies of physics reach of this detector. 

\begin{figure}[ht]
    \centering
    \includegraphics[width=14cm]{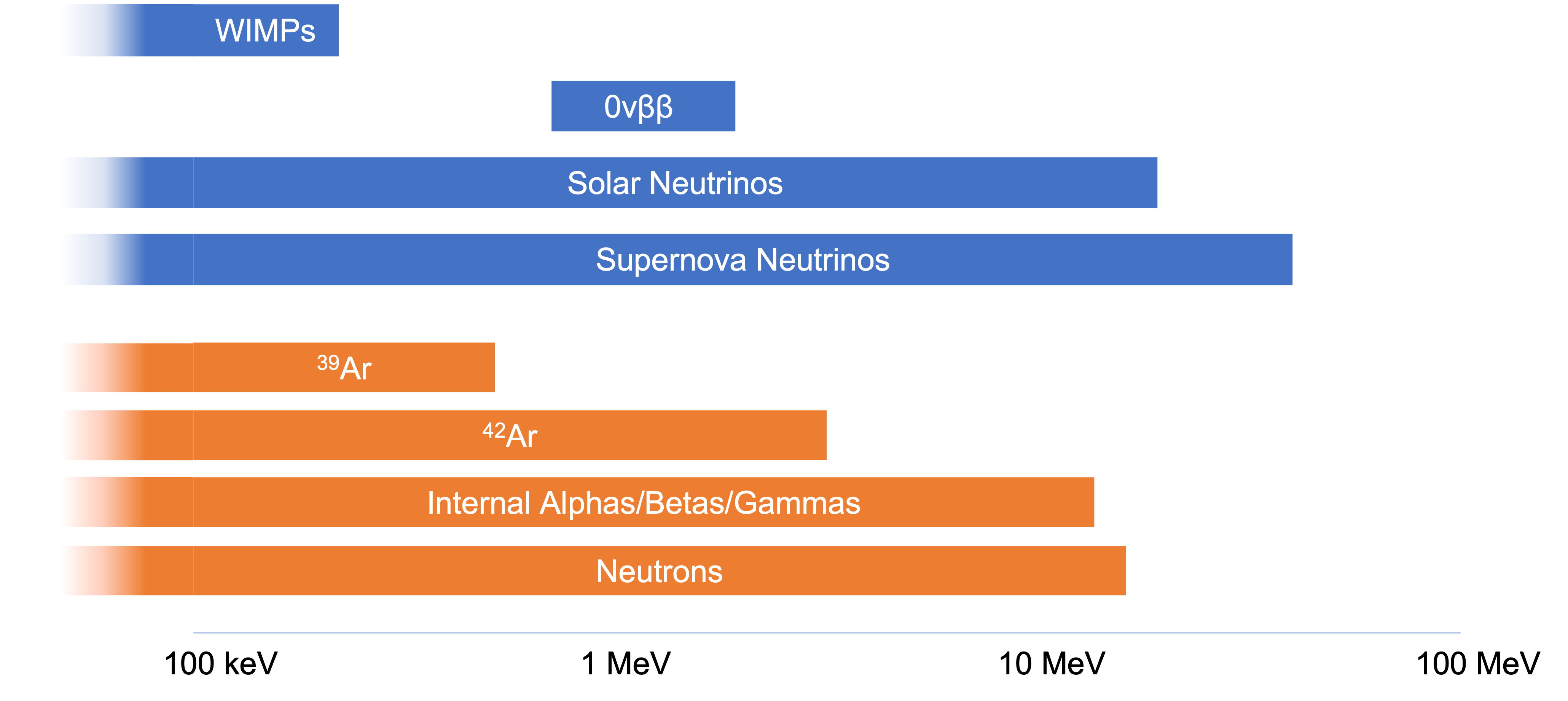}
    \caption{Summary of physics targets of this low background module and the primary radiological backgrounds.}
    \label{fig:summary}
\end{figure}

\section{Detector Design}\label{sec:DD}

This section outlines the proposed design for the low background module and describes in detail the radioactive background control and photon detection system enhancements required. We note this module design is not  endorsed by the DUNE collaboration.

\subsection{Module Layout}
A low background module is enabled most simply by minimizing the detector components in the bulk of the argon. The resulting module must still allow for very good light detection efficiency and for charge detection efficiency similar to existing designs. For the benefit of conforming to the longstanding plans for the DUNE far detector complex and cavern layouts, we also want to use the same commercial cryostat concept and existing module designs to the greatest extent possible and perturb them only where necessary. Starting with the existing module one or two designs assures minimal disruption to the main neutrino oscillation program.

\subsubsection{Single phase}
\begin{figure}[ht]
        \centering
        \includegraphics[width=\textwidth]{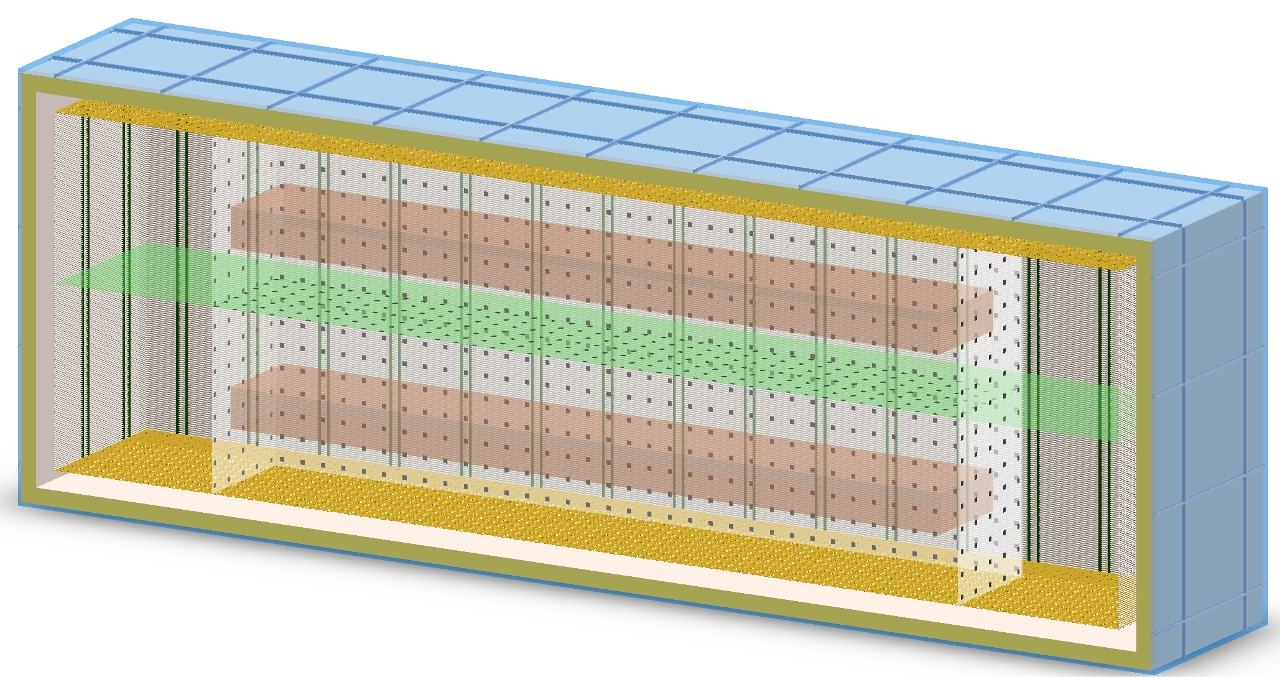}
        \caption{Shown is the base design for the proposed low background detector. Blue shows external water "bricks". The top and bottom yellow planes are the Charge Readout Panels unchanged from the Vertical Detector design. The central cathode is in green. The white box of acrylic (full interior volume) is of dimensions 6x12x20 (12x12x60) m$^3$. The black points are SiPM modules shown here at a low coverage of 10\% for viewing's sake, while some studies in this paper use up to 80\% coverage. A proposed fiducial volume totaling 2kT is shown in the two beige boxes. This paper also considers a 3 kT volume.}
        \label{fig:cadmodule}
    \end{figure}

We therefore start with DUNE's Module 2 Vertical Drift (VD) detector~\cite{snowmass_pof} and consider design modifications to suit our low background purposes. We show a working design in Figure~\ref{fig:cadmodule}. 

Water in "bricks" which are imagined to be nestled among the I-beam support structure achieve large external neutron reduction, as discussed. We keep the Charge Readout Planes of the VD unaltered. Similarly the central cathode of the VD is preserved with the exception that the sparse and mostly distant X-Arapucas are swapped out for the SiPM modules -- at least on that part of the cathode in the inner region of this detector. An acrylic box of one inch thickness with {x,y,z} dimensions {6,12,40} m serves to mount SiPM modules and reflective WLS foils. Here x is the horizontal dimension, y is the vertical dimension, z is in the beam direction. SiPM modules are mounted on the inside of the acrylic box at anywhere from 20-80\% area coverage. We envision two 1 or 1.5 kT  long skinny fiducial volumes, depending on the study, that have a stand-off of 1.5m from the central cathode. We also discuss, alternatively, a 3 kT fiducial volume in this paper in studies where backgrounds from the central cathode are thought to be small or events from it can be reconstructed and cut away.

The bulk volume of this module consists mostly of underground argon, with only small material contributions such as the slender support structures for the cathode plane panels. As in the VD, there are two 6m vertical drifts.

    
\begin{figure}[ht]
    \centering
    \subfigure[]
    {
\includegraphics[width=8cm]{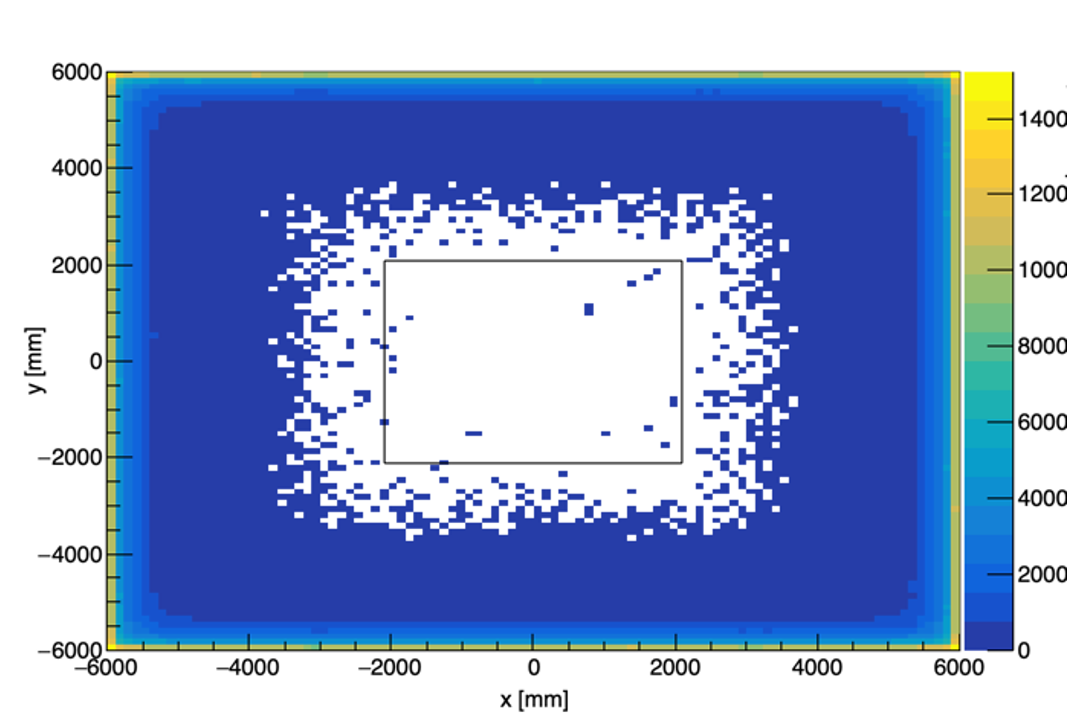}
    }
    \subfigure[]
    {
\includegraphics[width=8cm]{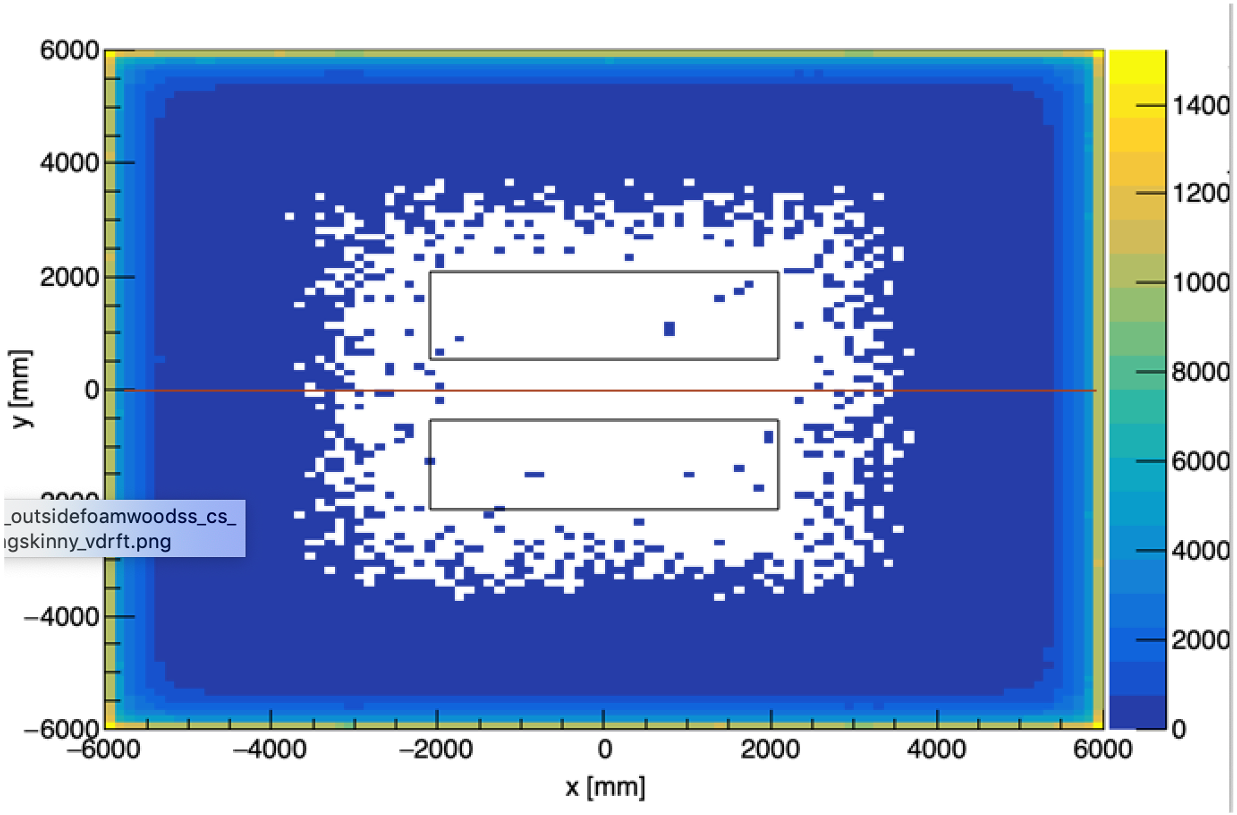}
    }
\caption{(a) Neutron interactions above 100 keV from stainless steel 15 ktyr exposure} for a simple rectangular parallelepiped fiducial volume and (b) a smaller total volume over two parallelepipeds that stand off from the central cathode.
\end{figure}

\subsection{Radioactive Backgrounds}

To enable the physics targets for this module, improvements in control of internal and external radioactive background levels are required. Making a module of this size low background will require significant quality and materials controls beyond what has been been attempted by previous experiments, and certainly beyond what is required for the baseline DUNE program. However, we note that due to the increased size of this module self-shielding in the argon allows the background requirements to be less stringent than those expected to be reached by the coming generation 2 dark matter experiments. Thus future research and development will need to focus on how to scale the techniques successfully deployed to low background dark matter or neutrinoless double beta decay searches to the kton scale.

A particular concern for this module will be neutron-induced background events, which will be the main background to the neutrino searches above 3.5~MeV thresholds. A neutron capture in $^{40}$Ar produces a 6.1~MeV gamma cascade which can Compton scatter or pair produce electrons which can mimic the charged current neutrino interactions in the argon. Captures on $^{36}$Ar can produce 8.8~MeV gamma cascades. Neutrons are also the primary backgrounds for the lowest energy searches for WIMP dark matter, where nuclear recoils can mimic the signal. Below 3.5 MeV the primary backgrounds will come from alpha, beta and gamma emitting isotopes within the argon or detector materials.

\subsubsection{Cavern Neutron Backgrounds}

\begin{figure}[ht]
        \centering
        \includegraphics[width=10cm]{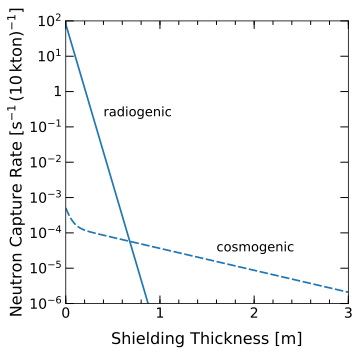}
        \caption{From~\cite{PhysRevC.99.055810}. The neutron capture rate reduction in liquid argon as a function of water shield thickness is shown.}
        \label{fig:watershield}
    \end{figure}

The most significant source of neutrons will likely be those induced by spontaneous fission or ($\alpha$, n) interactions from the uranium-238 or thorium-232 decay chains within the surrounding rock and shotcrete of the detector cavern. As reported in~\cite{BEST20161}, the external neutron rate at SURF (a likely hosting laboratory for this low background module) is assumed to be $1.0 \times 10^{-5}$~n/cm$^{2}$/s. As proposed in References~\cite{PhysRevC.99.055810}\cite{PhysRevLett.123.131803}, it is possible to add water shielding to a DUNE-like cryostat, taking advantage of the space between the structural supports even when the detector is located within a cavern at SURF with limited space around the detector. Figure~\ref{fig:watershield} from that reference shows that a 40~cm water shield, located within the support structure, is enough to lower the external neutron rate by three orders of magnitude. We assume this is achievable for this module.

\subsubsection{Cryostat Backgrounds}

The strategy to control the radioactive backgrounds from the detector components such as the cryostat will have three parts:
\begin{itemize}
    \item improvements to material selection;
    \item additional internal neutron shielding; 
    \item and advanced event selections and analysis tools,
\end{itemize}
with the aim of lowering the internal neutron rate within the detector by at least three orders of magnitude to match the levels of the external rates.

Aside from external neutrons from the cavern, the main source of neutrons in a DUNE-like detector cryostat is likely to be the order 1~kton of stainless steel that makes up the I-beam support structure. Research and development will be required to lower the internal background rates by the three orders of magnitude required, for example by careful selection of the raw ingredients and/or control of the manufacturing process. It should be noted that the ``Generation 2'' dark matter experiments expect to reach neutron rates from their steel a further two orders of magnitude beyond this, so the goal is achievable (even if the scale is much larger than previously attempted). Another area of R\&D required to support this goal is improvements in knowledge of ($\alpha$, n) cross sections, as highlighted in the recent IAEA workshop~\cite{JReichenbacherIAEA}.

Another approach is to improve the internal shielding within the detector by adding higher density R-PUF insulating foam and/or boron, lithium or gadolinium loaded material layers within the membrane cryostat structure. Our studies show that this could easily reduce the neutron capture rate in LAr by one order of magnitude. One approach, as planned by the DarkSide collaboration, would be to use the additional planes of Gd-doped acrylic to act as a neutron absorber. DarkSide-20k intends to use multiple layers within a ProtoDUNE-style cryostat for their dark matter search. Another design choice might be to take advantage of the existing cryostat but replace materials such as the insulating foam itself to reduce backgrounds from the support structure.

Analysis based cuts can also be used to remove events. For example with the low threshold of this planned detector neutron induced multiple scatters could be tagged and rejected. This takes advantage of the excellent ($\sim20$~mm) transverse resolution of a TPC. Studies to identify dark matter nuclear recoil backgrounds show $\sim30$\% reduction at 100 keV threshold and $\sim90$\% reduction at 50 keV.

\subsubsection{Radon and other internal argon backgrounds}
%

For this low background module we set a radon target level in the liquid argon of 2 $\mu$Bq/kg. This is about three orders of magnitude below the expected DUNE radon level of 1 mBq/kg~\cite{dune_bkgmodel1}\cite{dune_bkgmodel2}\cite{dune_bkgmodel3}. This level has been achieved in liquid argon by the DarkSide-50 experiment~\cite{DarkSide-20k}, and exceeded by DEAP-3600 which achieved a level of 0.2 $\mu$Bq/kg~\cite{PhysRevD.100.022004}. 

To reach this level in a kton-scale detector will require research and development to implement a combination of the following techniques:
    \begin{itemize}
        \item \textbf{Radon removal during purification via an inline radon trap}. No radon removal is in the current design of the purification system for the baseline DUNE. Dark matter and neutrinoless double beta decay search experiments typically use cooled, activated charcoal radon traps to remove radon directly from the recirculating target. The most sensitive dark matter experiments typically purify the argon in the gaseous phase~\cite{radon_xenon100}~\cite{PUSHKIN2018267}, however such an approach would be impractical for a kton-scale experiment. Borexino used a charcoal radon trap to purify liquid nitrogen~\cite{HEUSSER2000691}, and such an approach could be adopted and scaled appropriately for a low background module. New materials such as Metal-organic frameworks could improve the capture-potential beyond charcoal, allowing a potential shrinkage of footprint of a radon-capture facility to fit the existing cavern designs.
        \item \textbf{Emanation measurement materials campaign}. All materials used in detector construction are known to emanate radon at some level. A large-scale emanation assay campaign to identify materials suitable for construction, similar to the QA/QC campaign described above will be required to ensure the detector can meet the target. A topic for R\&D will be how to increase throughput of samples, as emanation measurements typically take two weeks per sample.
        \item \textbf{Surface treatments}. For large components such as the cryostat where it may be impractical and costly to make significant improvements to the radiopurity, surface treatments can be used to lower emanation rates. It is known that acid leaching and electropolishing lowers emanation rates. Coating the inner surface of the cryostat with a radon barrier could lower emanation rates from this significant source.
        \item \textbf{Dust control}. Dust is a significant radon source and cleanliness standards will be higher in this low background module than the baseline DUNE design. Cleanliness protocols and requirements will be required and R \& D will be required to develop automated techniques of use to, for example, the several square km's of surface area of a DUNE-like cryostat.
        \item \textbf{Radon reduction system during installation and operation}: The mine air underground is radon laden up to 1,000 Bq/m$^{3}$ and radon daughter plate-out during installation, filling and operation must be controlled. An upscaled vacuum swing system with large charcoal columns in parallel to remove radon from the ambient air and to provide radon-free air to the cleanroom and cryostat would be suitable. Vacuum swing systems providing radon-free air have been successfully employed by e.g. Borexino~\cite{doi:10.1063/1.2060466}, LZ~\cite{LZ_bkgs} and SuperCDMS~\cite{doi:10.1063/1.5018995}. 
        \item \textbf{Drifting of charged daughters to cathode}. Several daughters in the radon chain are charged and will drift towards the cathode and out of the fiducial volume. This effect may be countered by mixing effects of the purification system however.
        \item \textbf{Alpha tagging through pulse shape discrimination}. Alpha events in the radon chain that produce neutrons directly in the argon may be taggable, by identifying the alpha track before the ($\alpha$,n) event. Though the amount of light may be relatively small, the timing profile is distinct and may allow pulse shape discrimination on this module with enhanced optical systems.
    \end{itemize}

\subsubsection{Underground Argon}
\label{sec:UAr}
\begin{figure}[ht]
        \label{DarksideUAr}
        \centering
        \includegraphics[width=12cm]{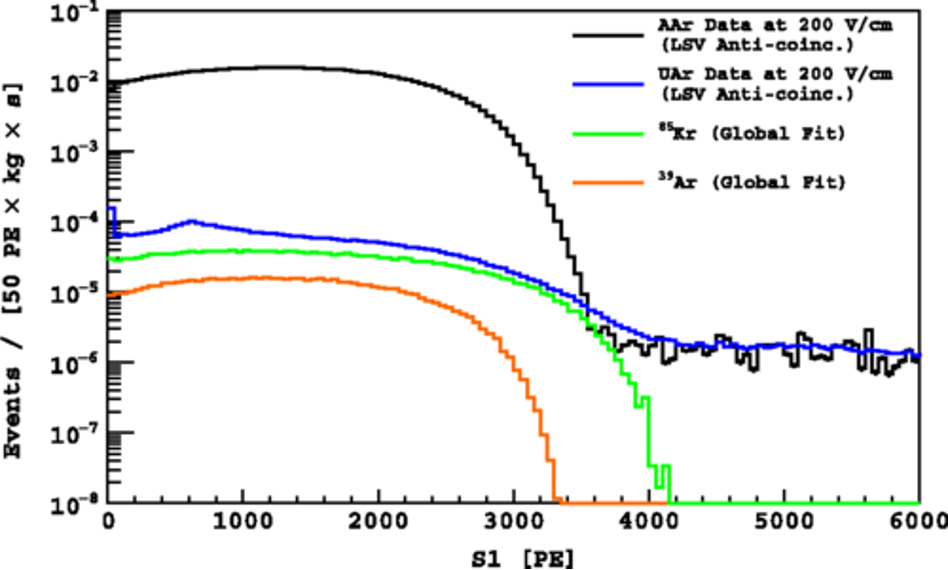}
        \caption{$^{39}$Ar and $^{85}$Kr from DarkSide-50 showing a x1400 reduction in $^{39}$Ar for UAr. \cite{UAr_DS50_39Ar}}
        \label{fig:uar}
    \end{figure}

Natural argon is mostly $^{40}$Ar which is not radioactive. There are some long-lived radioactive isotopes-$^{39}$Ar ($T_{1/2}$=269y, $Q_{\beta}$=565 keV), $^{37}$Ar ($T_{1/2}$=35d, $Q$=813 keV), $^{42}$Ar ($T_{1/2}$=32.9y, $Q_{\beta}$=599 keV) \cite{nndc} which are , in atmosphere, produced primarily by cosmic ray-induced reactions in $^{40}$Ar. The use of atmospheric argon in a multi-ton scale argon detector has limitations due to high $^{39}$Ar activity (1 Bq per kg of argon \cite{WARP_2007}) in atmospheric argon (AAr). Radiogenic and cosmic-ray muon-induced interactions, especially on K and Ca isotopes, can produce $^{39}$Ar underground. The dominant production channels are negative muon capture on $^{39}$K and  ($\alpha,n$)-induced (n,p) reactions on $^{39}$K. $^{39}$Ar production underground decreases significantly with depths\cite{Mei_2010} as $-ve$ muon flux decreases with depths. DarkSide-50, the only experiment to use underground argon (UAr), measured the $^{39}$Ar activity of 0.73 mBq/kg \cite{UAr_DS50_39Ar}, a factor of 1400 smaller than in AAr (see Figure~\ref{DarksideUAr}). With the ARIA project\cite{nowak2021separating}, the DarkSide collaboration is planning to deplete UAr of $^{39}$Ar through large-scale isotopic separation by cryogenic distillation.

In the atmosphere, the $^{42}$Ar concentration is $\sim$ $10^{-20}$ $^{42}$Ar per $^{40}$Ar atom {\cite{Barabash_2016},\cite{PEURRUNG1997425}}, which is four orders of magnitude smaller than $^{39}$Ar's concentration. The daughter isotope of $^{42}$Ar, $^{42}$K ($T_{1/2}$= 12 h) has two major decay modes : i) Direct beta-decay to the ground state of $^{42}$Ca ($Q_{\beta}$= 3525 keV, BR=81 $\%$), and ii) Beta-decay ($Q_{\beta}$= 2001 keV) to an excited state of $^{42}$Ca followed by a prompt 1524 keV gamma emission. In the atmosphere,$^{42}$Ar can be produced primarily by two reaction channels. One is two-step neutron capture: $^{40}$Ar(n,$\gamma$)$^{41}$Ar followed by $^{41}$Ar(n,$\gamma$)$^{42}$Ar. Since the intermediate isotope $^{41}$Ar is short-lived, any significant production of $^{42}$Ar through this channel requires high neutron fluxes like that generated in nuclear tests/explosions \cite{cenn1995}. The other is $^{40}$Ar($\alpha$,2p)$^{42}$Ar, primarily occurring on the upper atmosphere \cite{PEURRUNG1997425}, where energetic alphas are readily available from cosmic ray muon interactions. The $^{42}$Ar production rate underground is not known, but it is expected to be several orders of magnitude smaller than in AAr.  Particle interactions on isotopes of K, Ca, and Ti can produce some $^{42}$Ar in the earth's crust, given the relatively high abundance of the elements (by mass-fraction\cite{CRC_2021}:K-2.09$\%$,Ca-4.15$\%$,Ti-0.565$\%$). However, the reaction thresholds are high,  making $^{42}$Ar production energetically not possible by fission, ($\alpha$,n)-neutrons and alphas from the natural radioactivity chains of $^{238}$U, $^{235}$U, and $^{232}$Th. Energetic particles from cosmic ray muon-induced interactions can produce $^{42}$Ar. But at the depths at which underground argon is usually extracted, the cosmic ray muon-flux should be hugely suppressed, so $^{42}$Ar production should be negligible. $^{42}$Ar decays in the bulk  argon volume will produce $^{42}$K isotopes. While the $^{42}$Ar beta-spectrum has an endpoint of 599 keV, Betas from $^{42}$K-decays span a much larger energy range up to 3.5 MeV, and can be problematic backgrounds. Since UAr should be heavily depleted of $^{42}$Ar, significant suppression of $^{42}$K decay backgrounds is achievable with UAr. Based on GERDA's findings\cite{lubashevskiy2018mitigation}, following $^{42}$Ar decays, $^{42}$K nuclei could retain the positive charge long enough to drift across in the influence of electric field. So, we expect $^{42}$K ions to drift and move towards the cathode plane- which suggests an additional suppression of $^{42}$K backgrounds is achievable through fiducialisation. 

The $^{85}$Kr isotope, predominantly, a $\beta$-emitter has a half-life of 10.7 years and Q-value of 687 keV\cite{nndc}.  Primary modes of $^{85}$Kr production are spontaneous fission of uranium and plutonium isotopes, neutron capture on $^{84}$Kr, and human-induced nuclear fissions in nuclear reactors\cite{schroder1971physical}. We would expect $^{84}$Kr to be present at some level in AAr, however as a short-lived isotope its concentration can vary across argon extraction sites, most likely being higher in the proximity of nuclear sites.  Using the UAr data, DarkSide-50 measured $^{85}$Kr activity of 2 mBq/kg\cite{UAr_DS50_39Ar} a few orders of magnitude smaller than in AAr. $^{85}$Kr concentration in UAr should also vary depending on the location of the gas reservoir and gas origin (mantle-like or crustal-like). Pacific Northwest National Laboratory (PNNL) will assess the production of $^{85}$Kr  in the underground gas wells that are potential sources for UAr. 

 We expect argon gas extracted from an underground source to be highly depleted of isotopes -$^{39}$Ar, $^{42}$Ar, $^{85}$Kr. There is  evidence that air infiltration during the UAr extraction could have contributed to the DarkSide-50's $^{39}$Ar - actual $^{39}$Ar content in the UAr could be significantly smaller (on the order of few tens of $\mu$Bq/kg)\cite{andrew_renshaw_2018_1239080}. $^{85}$Kr and $^{42}$Ar content could also be much smaller. Unlike stable gas isotopes like $^{40}$Ar, which can get collected at gas wells over time, short-lived heavy isotopes like $^{39}$Ar($T_{1/2}$=269y), $^{42}$Ar ($T_{1/2}$=32.9y) and $^{85}$Kr($T_{1/2}$=10.7y) diffusing through rocks and getting collected in a significant number at the underground gas wells is less likely. However, air-infiltration and cosmogenic activation in the argon bulk could introduce these isotopes in the extracted UAr~\cite{saldanha39-37,zhang2022evaluation}. Greater care, perhaps, is necessary to ensure avoiding contamination of the UAr during extraction, processing, transport and storage.

While UAr is desirable, it requires a dedicated effort to identify the potential argon source and procure argon on a large enough scale necessary for this project. The Urania plant~\cite{aalseth2018darkside} in Southwestern Colorado is expected to produce underground argon from CO$_2$ gas wells $\sim$ 300 kg/day (at full rate) for DarkSide. The argon source and the production rate is not large enough for a kilo-ton scale experiment. The authors are in the process of identifying alternative gas wells with enriched argon streams. PNNL is already discussing with three potential commercial suppliers, however the underground source samples are not yet tested and low levels of radioactive isotopes must be proven. Initial gas analysis indicates the mantle origin of these sample. Based on estimates by the gas suppliers, the production cost could be as low as 3x cost of atmospheric argon and 5kTon of argon production per year could be achievable.  

\subsubsection{Surface storage and spallation issues}

During above-ground storage of our UAr, before placement into the underground module, the isotopes $^{37,39}$Ar are produced by cosmogenic neutrons in the reactions $^{40}$Ar(n,4n)$^{37}$Ar, $^{36}$Ar(n,$\gamma$)$^{37}$Ar or $^{38}$Ar(n,2n)$^{37}$Ar and $^{39}$Ar in reactions: $^{40}$Ar(n,2n)$^{39}$Ar and $^{38}$Ar(n,$\gamma$)$^{39}$Ar reactions.  $^{42}$Ar, can be produced in the two steps reactions $^{40}$Ar(n,$\gamma$)$^{41}$Ar and thus $^{41}$Ar(n,$\gamma$)$^{42}$Ar. In a previous work \cite{Parvu_2021}, to which we refer the interested reader, the cross sections for these reactions were obtained from nuclear reaction codes and confronted with data where they exist. A full study of expected spallation and pileup backgrounds during the detector operation is in reference~\cite{PhysRevC.99.055810}.

\subsection{Light Collection Enhancement}

The light collection for this low background module will be enhanced to enable two main goals:
\begin{itemize}
    \item lower the energy threshold and improve the resolution at these lower neutrino energies;
    \item improve pulse shape discrimination for radioactive background rejection.
\end{itemize}
In this section we describe the improvements to the light detection system required to enable this.

\subsubsection{Photosensors}
\label{photosensors}

As a baseline design our plan is to use Darkside-20k style SiPM tiles~\cite{aalseth2018darkside} (24 cm$^2$) with 50\% quantum efficiency. We assume here another 50\% WLS efficiency from, e.g., TPB on the tile surface, for a total efficiency of 25\%. For maximum light detection we envision covering both the central cathode's hollow middle (as will be done in module 2) and the interior acrylic walls at up to 80\% coverage. We assume the module 2 power-over-fiber concerns~\cite{snowmass_pof} to be solved to allow our SiPM coverage of the cathode. The resulting number of SiPM modules for, as an example, 10\% coverage -- a value which current studies naively show is sufficient, but which needs more detailed work -- is $\sim 50000$, to be compared to Darkside20k's planned $\sim 10000$. It is likely possible to optimize the placement of the tiles around the fiducial volume, to reduce the total number required, and work is ongoing to study this.

\subsubsection{Reflectors}

To maximize the light capture in this module the surface of the acrylic box will be coated with a reflector to create a light-tight inner volume. If a PTFE reflector is used, as in DarkSide-50, reflectivities of ~97\% should be possible.

\subsubsection{Argon Purity}

The baseline requirement for DUNE is $<25$~ppm of nitrogen to ensure that photon propagation is not quenched. For our simulation studies we assume that attenuation (absorption) lengths of order 50~m are achievable, which corresponds to nitrogen contamination of 1-2 ppm~\cite{Jones_2013}. We note that dedicated dark matter experiments have achieved ppb levels of purity. 

\subsubsection{Light Collection Efficiency Impacts on Energy Resolution}
\label{enressection}

Charged particles that traverse the liquid argon deposit energy and excite and ionize the argon atoms. This process results in the electrons recombining with the ions generating unstable argon dimers, which decay and emit scintillation light. If an electric field is applied, a fraction of the electrons will drift away before recombining. These ionization electrons are collected by the anode plane and create the charge signal by removing electrons that would have recombined yields an anticorrelation between the light and charge signals observed in LArTPCs. The small number of ionized argon atoms at low energies means that fluctuations in this recombination process can smear the amount of charge observed to energy deposits. The Noble Element Simulation Technique (NEST) collaboration has explored the precision of LArTPC for 1 MeV electrons as a function of charge readout signal-to-noise ratio and the efficiency of collecting light~\cite{nesteres}. They predict that using only the charge signal in a LArTPC, the most precise one can reconstruct the energy for a 1 MeV electron is 5\% and MicroBooNE has validated this prediction~\cite{bhat_phd}. To improve the energy reconstruction, further one needs to include measurements of scintillation light. 

\begin{figure}[ht]
    \centering
    \includegraphics[width=0.55\textwidth]{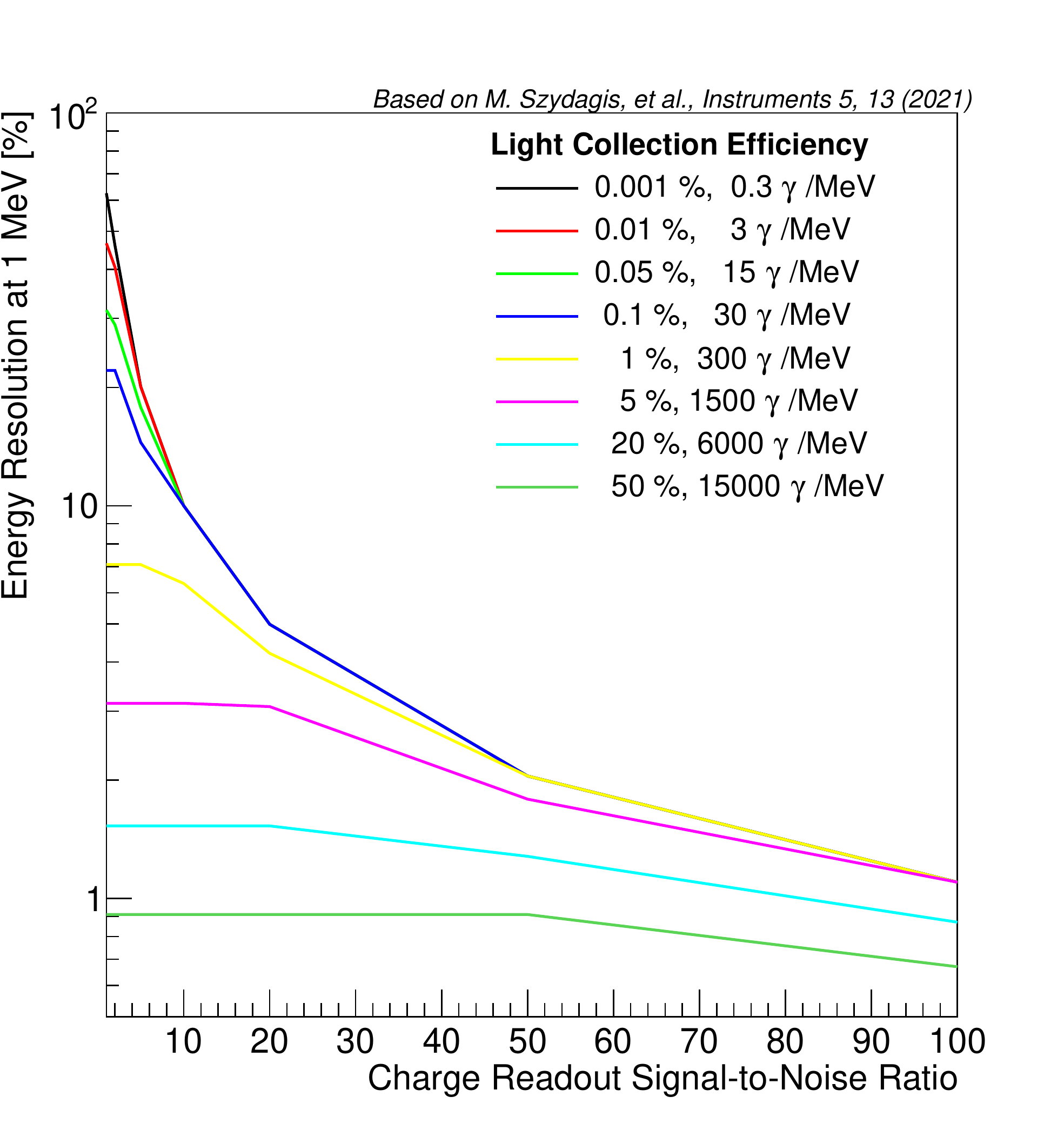}
    \caption{Data from Reference~\cite{nesteres} reinterpreted. The best energy resolution achievable by a LArTPC for a 1~MeV electron plotted versus charge readout’s signal-to-noise ratio (x-axis) and the detector’s efficiency for collected scintillation light from the electron (colored lines). This figure is from Reference~\cite{dunebeta}.}
    \label{fig:EresvsLight}
\end{figure}

The NEST collaboration explored the expected energy resolution enhancements that a LArTPC can achieve by including light signals. Fig.~\ref{fig:EresvsLight} encapsulates this information by showing the energy resolution for a 1 MeV electron for LArTPCs with different signal-to-noise ratios (SNR) and varying light collection efficiency. LArTPCs with SNR near 40 need 50\% of the scintillation light to measure energy deposits with 1\% precision. 

\section{Physics Studies}\label{sec:phys}

This section outlines key physics goals of this module and describes the studies that have been performed.

\subsection{39Ar Studies}

Reduction of the rate of $^{39}$Ar is desirable for a number of reasons. This 600 keV endpoint beta emitter decays at a 1 Bq/L rate in ordinary atmospheric argon. It may mask low energy physics by creating optical signals that confuse the reconstruction process. It also represents a serious hurdle in the detector triggering considerations.

Trigger primitives, which constitute a 6 PByte/year data source, not necessarily to be stored into perpetuity, is still an onerous data flow to deal with during steady data-taking. That number drops to a far more manageable 10s of TBytes if the $^{39}$Ar is reduced by a factor of 1400. In reality of course, much of the data budget will perhaps be consumed by low-threshold activity in this detector. 

Perhaps the more pernicious practical problem is that $^{39}$Ar  beta decays may reconstruct as optical Hits which may, in turn, comprise Flashes which then confuse the charge-light association for reconstructed physics objects -- especially at low energy and far from the light detectors. We have performed a study in a Module 1-like environment, not shown here, that gives the not very surprising conclusion that Supernova flashes become unambiguously matched with a x1400 $^{39}$Ar reduction.

Third, the highly desirable property of Pulse Shape Discrimination (PSD) in Argon which would allow to subtract out the nuisance $^{39}$Ar contribution for our, e.g., Dark Matter search ambitions, is not practicable if pile-up is too intense. We show in~\cite{DUNE-DM-PNNL} and discuss later in this paper that a x1400 reduction of  $^{39}$Ar just allows for a search in our fiducial volume down to 100 keV thresholds and perhaps lower.

Here we  mention that even a x1400 reduction of  $^{39}$Ar does not allow this detector to get to arbitrarily low thresholds for searches of physics with electronic signatures -- as apposed to neutron-like interactions. We expect reductions by this amount still leave $^{39}$Ar as an overwhelming background to low-rate, low-energy solar processes, for example.

\subsection{Simulation}
\label{simulation}

Almost all studies in this paper are carried out in a standalone Geant4\cite{geant4} simulation. Source code and build instructions are found at Reference~\cite{rdk}. In that simulation is proper isotope decay and neutron physics, along with optical physics. The volume is basically a 10 kT box of liquid argon, but with a reasonable model of the cryostat walls on all six sides with charge readout planes (CRPs) made of G10 on the floor and ceiling, as in the VD. Further, there is an acrylic box inside, open on top and bottom and tiled with 24 cm$^2$ SiPM modules at an 80\% coverage. SiPM modules with that same coverage viewing both upper and lower volumes also tile the central cathode plane. See Figure~\ref{fig:cadmodule} for a faithful representation of our simulated geometry. We use an after-the-fact 25\% total quantum efficiency. There is no SiPM electronics response applied. We include a 96\% reflectivity of the acrylic surface and a 44\% reflectivity for the CRPs (as the holes in each CRP are about 56\% of the surface area), and impose an argon attenuation (absorption) length of 50m, and a Rayleigh scattering length of 90cm. All simulations generate their 128 nm photons ab initio from the charge particles which create them and are propagated on the fly, with no lookup libraries, until their end point on a SiPM where they are counted or their disappearance through absorption.

\subsection{Optics Studies}
\label{optics}

The simulation described in Section~\ref{simulation} was used to study the minimal requirements for the optical system to allow pulse shape discrimination in a dark matter search, including the required reflectivity of the acrylic box and anode readout surfaces as a function of SiPM tile coverage. The minimal photon counting requirement for the optical system was set at 400 photons reaching the SiPM surface, which results in a total of 100 photons being detected due to the assumed 25\% efficiency described in Section~\ref{photosensors}. The 400 (100) photon requirement was chosed as the minimal amount of photons required to perform a pulse shape discrimination analysis such as described in Section~\ref{sec:darkmatter}.

\begin{figure}[htpb]
    \centering
    \subfigure[]
    {
\includegraphics[width=12cm]{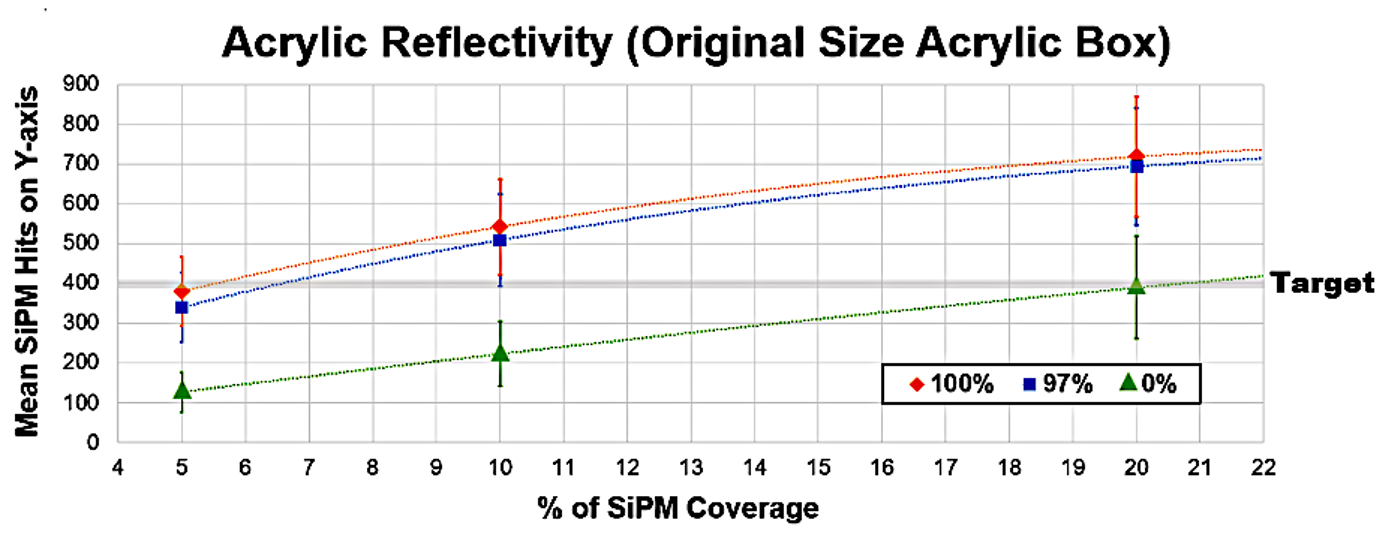}
    }
    \subfigure[]
    {
\includegraphics[width=12cm]{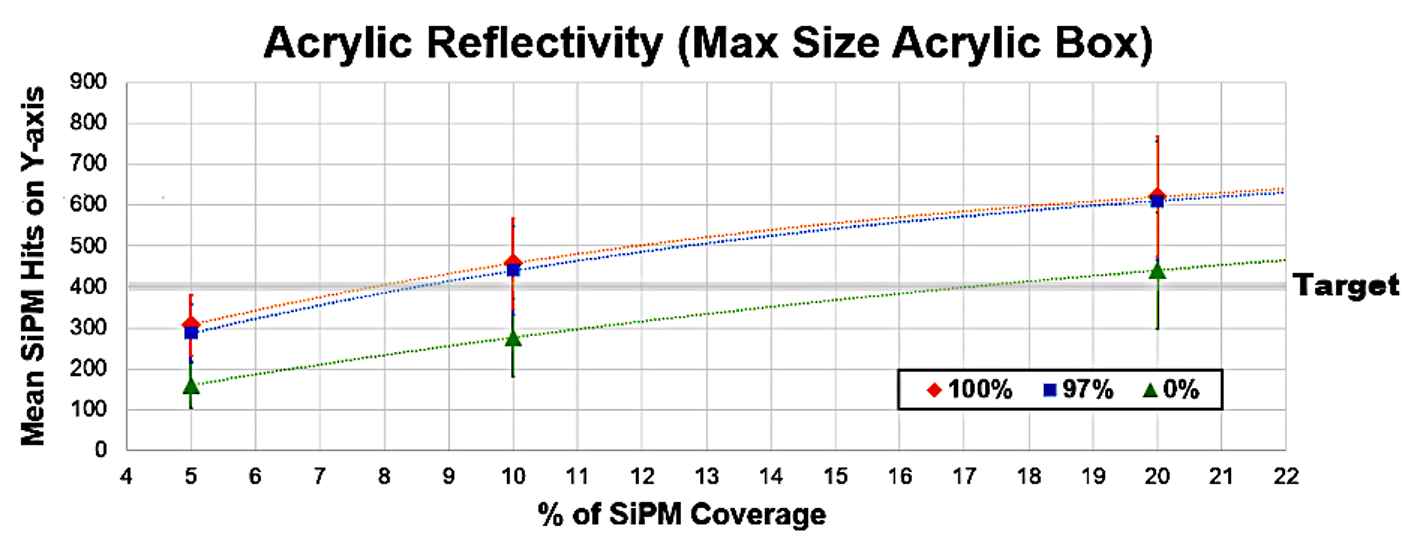}
    }
    \subfigure[]
    {
\includegraphics[width=12cm]{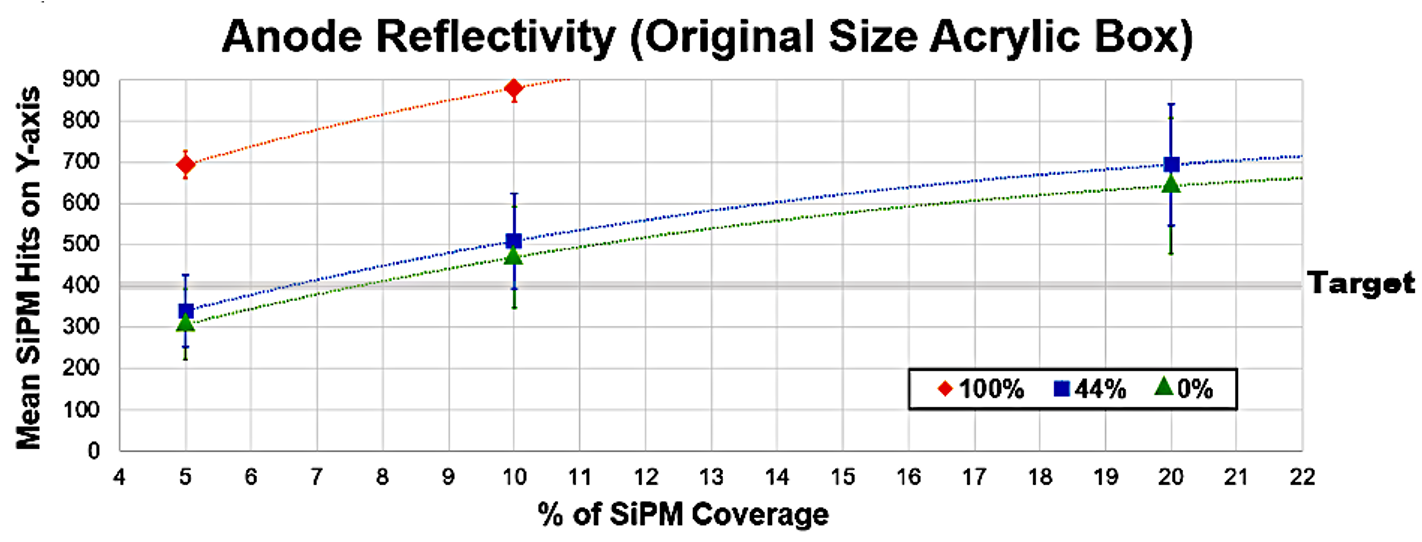}
    }
    \subfigure[]
    {
\includegraphics[width=12cm]{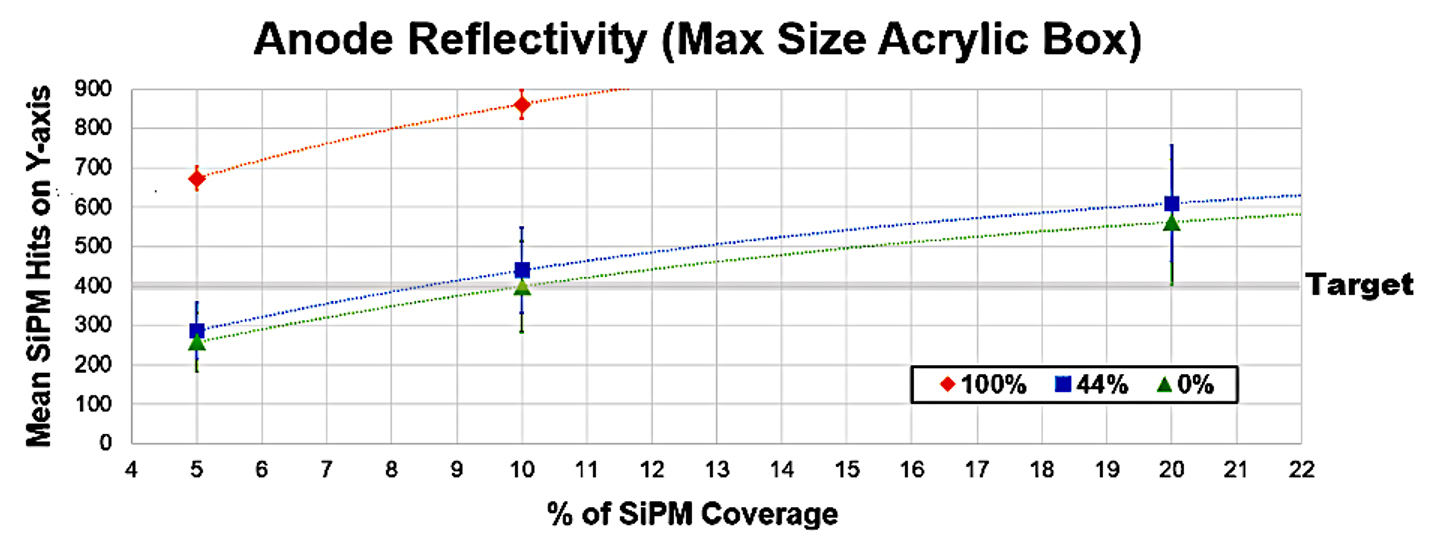}
    }
\caption{Number of photons detected at the SiPMs as a function of coverage varying both reflectivity of the surrounding acrylic and anode plane walls, and the size of the acrylic box containing the inner volume.}
\label{fig:reflect}
\end{figure}

The results of the simulation are shown in Figure~\ref{fig:reflect}. The studies were performed varying acrylic and anode reflectivity between 0~\% and 100~\% (a realistic 97~\% is highlighted) and then varying the SiPM coverage on the walls and cathode plane to count the accepted photons. Both the standard acrylic box described above and also a maximally sized box where the walls are moved out to the edges of the detector were simulated. This large box would be the worse-case optics scenario, maximising the path length of the photons. The studies show that a relatively modest amount of SiPM coverage of 10-20~\% is required even in the worst case scenarios to reach the target.

\begin{figure}[ht]
        \centering
        \includegraphics[width=12cm]{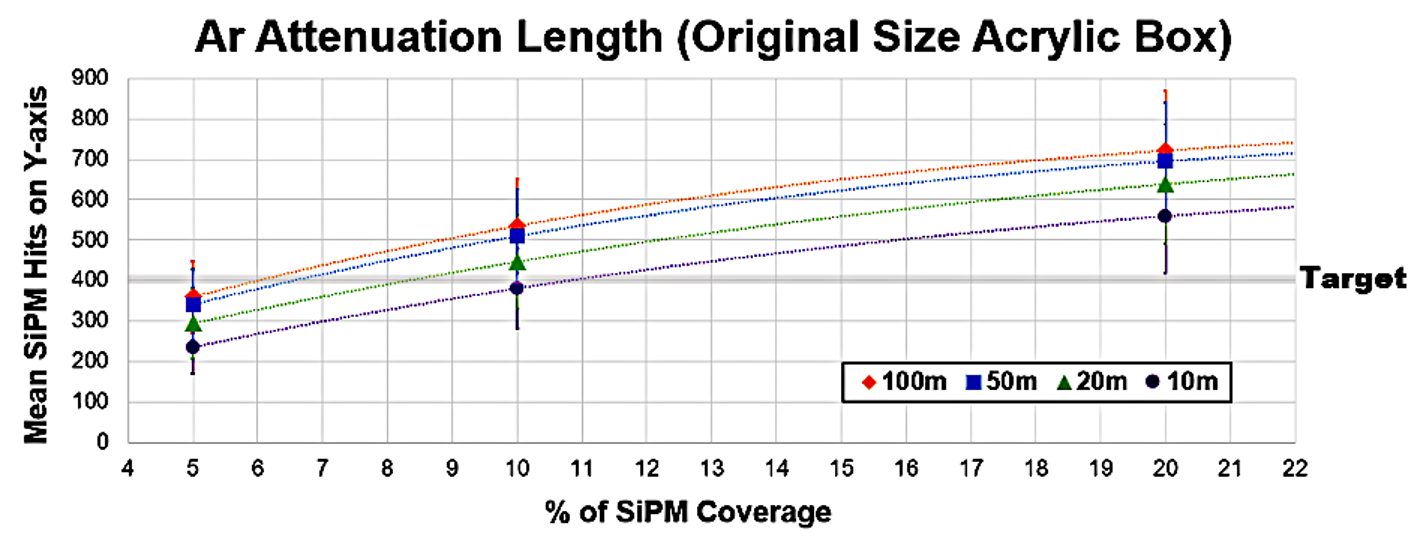}
        \caption{Number of photons detected as a function of SiPM coverage when varying the attenuation length within the argon.}
        \label{fig:atten}
    \end{figure}

The effect of attenuation with the liquid argon was also studied. Figure~\ref{fig:atten} shows the number of photons detected at the SiPMs as a function of SiPM coverage for a variety of different attenuation lengths. Assuming the relation between nitrogen contamination of the argon versus attenuation found in~\cite{Jones_2013}, this study shows that the 10-20\% SiPM coverage is sufficient to tolerate 0.5-5 ppm levels of nitrogen within the argon.

\subsection{Supernova Neutrino Physics}

In this section we present several supernova neutrino burst studies that could be enhanced by this module. This includes increased to sensitivity to lower energy, later time and source from greater distances. We also discuss the CE$\nu$NS glow sensitivity. 

\subsubsection{Supernova Energy Spectrum}

We wish to explore the potential improvement provided by a low background large LArTPC to that achievable in current DUNE Far Detector designs for sensitivity to supernova explosion detection. To this end we simulate a ten second exposure of our detector to a flux of electron neutrinos from the  Livermore~\cite{livermore} model and run them through the MARLEY~\cite{marley} generator to produce the final state particles in liquid argon. The detector response is provided by the simulation described in previous section~\ref{simulation}.

\begin{figure}[ht]
        \centering
        \includegraphics[width=\textwidth]{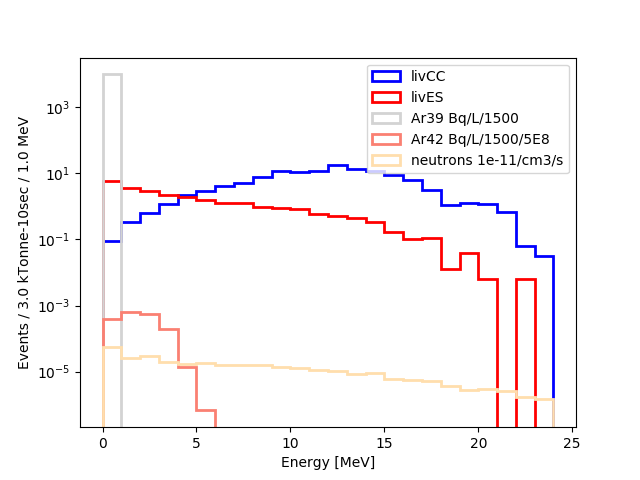}
        \caption{Neutrinos from supernova located at 10 kpc, assuming the Livermore model, during a ten second detector exposure window. Here we presume no Rn222 contribution and a likely too conservative x500 Ar$^{42}$ suppression in UAr with respect to that achievable in atmospheric argon. Reaching 5 MeV SN detection is straightforward in this module.}
        \label{fig:snnu}
    \end{figure}

Using that simulation we  count SiPM hits for an 80\% coverage and convert to energy using a simple conversion. That conversion comes from running 3 MeV electrons uniformly through the detector and counting SiPM hits in a coarse x,y,z binning of the production point. So, certainly there's room for refinement and reason to think that SiPM energy resolution could be greatly improved. And merely adding the charge response in each event above, say, 2 MeV would greatly improve the energy resolution by itself. We reiterate, for this study, we're only counting hit SiPMs and converting to energy. Events must originate in the inner 3 kT fiducial volume. The result is shown in Figure~\ref{fig:snnu}. The most evident feature is that the elastic scattering component of the $\nu_e$ flux  becomes dominant at low energies inaccessible to the baseline DUNE far detectors -- and it does so because neutron rates are required to be low from the cold cryoskin stainless steel and the $^{42}$Ar is at very low levels in UAr. The threshold here can go all the way down to 600 keV, which is a factor of roughly 18 lower than the baseline DUNE far detector design.


\subsubsection{Supernova Pointing}
The low detector threshold allows access to a significant number of elastic scatter events within the liquid argon. This opens up the possibility of reconstructing the position of the supernova with a pointing analysis. Liquid argon TPCs, with the excellent track resolution, are well suited to making this measurement. With the expected reduction in backgrounds in this sample, a clean elastic scatter sample will dominate at thresholds below 5 MeV (see Figure~\ref{fig:snnu}).

\subsubsection{Supernova Trigger}


The trigger system in a DUNE-like LAr detector relies on the so-called Trigger Primitives (i.e. hits, TPs) generated from the electronics connected to the wires or photo-sensors. They are simple objects constructed from the signal waveform. 
%
%
%
A stream of TPs arrives in the trigger module of the DAQ, which will use them to form a Trigger Activity (i.e. a cluster of hits, TAs), which is an association in time and space done by a trigger algorithm. A TA is related to each sub-module/component of the detector (e.g. an APA module), and multiple TAs form a Trigger Candidate (TC). Having a positive TC, the readout system stores the requested data. For low energy physics, which does not have an external trigger like beam events, it is essential to understand the detector backgrounds (e.g. radiological and electronics noise), to avoid triggers issued by undesired data. Thus, a Low Background LAr detector can perform better than the current designs.

The bottleneck for designing efficient supernova neutrino burst (SNB) trigger algorithms is the data transfer and storage resources available for the detector. To get the most of an SNB event, around 100~seconds of full detector readout is desirable, which means about 150~TB of raw data for a 10~ktonnes LAr detector. It will take about one hour to transfer the data from this trigger event from the detector caverns to the storage on the surface and several additional hours to transmit those data to the primary storage centre. Therefore, while the effective threshold must be set low enough to satisfy the requirements on SNB detection efficiency, it is crucial to not fire too frequently on background fluctuations. A requirement on the fake trigger rate of once per month is determined by these limits on data-handling, for a DUNE-like LAr detector. Additionally, the triggering decision needs to be made within 10~seconds since this is the typical amount of data buffered in the DAQ.

\begin{figure}[hptb!]
    \centering
    \includegraphics[width=\textwidth]{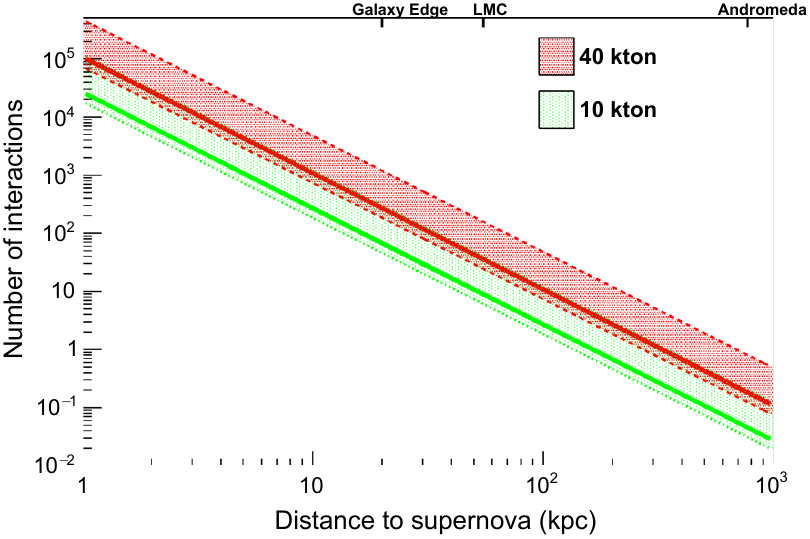}
    \caption{Estimated number of supernova neutrino interactions in a LAr detector as a function of the distance to the supernova, over a 10 seconds~\cite{DUNE:2020zfm}.}
    \label{fig:sn_interactions}
\end{figure}

\begin{figure}[hptb!]
    \centering
    \includegraphics[width=0.65\textwidth]{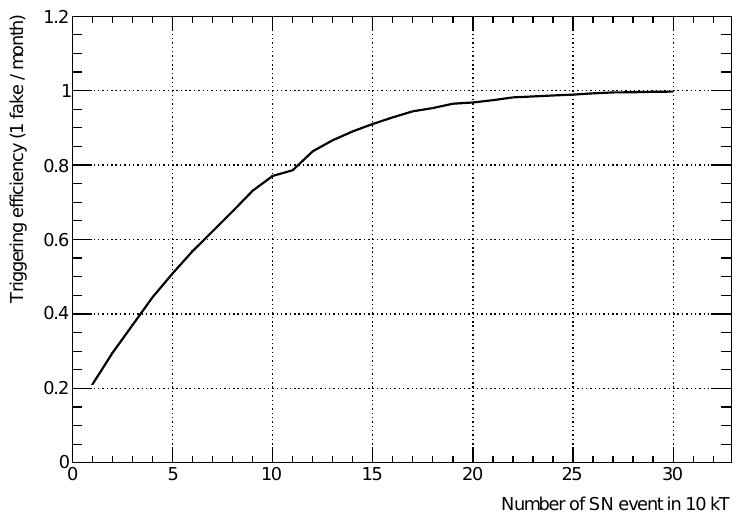}
    \caption{Supernova burst trigger efficiency as a function of the number of neutrino interactions, for an approach that utilizes the sum digitized-charge shape information of trigger activities input into the trigger decision~\cite{DUNE:2020zfm}.}
    \label{fig:sn_eff}
\end{figure}

Both TPC and PDS information gives a tagging efficiency of about 20-30\% for a single neutrino interaction~\cite{DUNE:2020zfm}. Reducing the estimated neutron capture rate in the LAr volume by a factor of ten (which is the principal background for SNB triggering, given the higher de-excitation gamma energy of about 6 MeV than other radiological components), this efficiency improves to 70\%, which translates to a 100\% (20\%) SNB triggering efficiency for a Milky Way (Magellanic clouds) SNB. The trigger strategy described above is a ``counting'' method. If we utilise the integrated charge of each TP, it is possible to construct a distribution of the TAs raw energy (SNB signal with backgrounds) and compare it to the background-only one. The ``shape'' triggering method improves the efficiency to tag Magellan clouds' SNB (or any SNB producing ten neutrino interactions, see Figure~\ref{fig:sn_interactions}) to 70\%. Figure~\ref{fig:sn_eff} shows the SNB efficiency as a function of the number of neutrino interactions in a ten ktonnes DUNE-like LAr detector using the standard DUNE background model described in Reference~\cite{abi2020deep} and a shape triggering algorithm that keeps the fake trigger rate to one per month.
%

With the Low Background LAr detector, a less stringent selection can be used, increasing the signal efficiency while keeping the fake trigger rate at the requirement level. Thus, higher efficiencies will be reached with a lower number of SN events in Figure~\ref{fig:sn_eff}, it then being possible to achieve 100\% efficiency to trigger a Magellan cloud SNB. Identifying Andromeda's SNBs is more challenging even with this design since they do not produce enough interactions in the whole detector volume in a 10 seconds window (see Figure~\ref{fig:sn_interactions}), and lowering the TA requirements would encounter the $^{39}$Ar activities.

\subsubsection{Late Time Supernova Neutrinos}

The neutrino flux from a core-collapse supernova is expected to cool over a few tens of seconds, with the late time events getting lower and lower in energy.  The tail end of the burst, where a black-hole-formation cutoff may be present, will be challenging to observe.  A lower energy threshold extends the time range with which a large liquid argon detector can follow the evolution of the supernova burst~\cite{Li:2020ujl}.

%


\subsubsection{Pre-supernova Neutrino Signal}

Another benefit of lowering the threshold is potential sensitivity to presupernova neutrinos~\cite{Odrzywolek:2003vn,Odrzywolek:2004em,Odrzywolek:2010zz,Kato:2017ehj,Patton:2015sqt,Patton:2017neq,Mukhopadhyay:2020ubs}. 
The final stages (hours to days) of stellar burning before the core collapse are expected to be associated with an uptick in neutrino production and energy, and observation of these could provide a true early warning of a core-collapse supernova.  The presupernova flux is expected to be small, and energies are typically less than 10 MeV; nevertheless they may be observable in a large liquid argon detector with low threshold~\cite{Kato:2017ehj} for  progenitors within a few kpc nearing the ends of their lives.

\subsubsection{CEvNS Glow}

Coherent elastic neutrino-nucleus
scattering (CEvNS)~\cite{cevns1},~\cite{cevns2} is a process that occurs when a neutrino interacts coherently with the total weak nuclear charge, causing the ground state nucleus to recoil elastically.  The cross section is large compared to the inelastic charged- and neutral-current interactions, but resulting nuclear recoil energies are in the few tens of keV range.  CEvNS has now been observed in argon by the COHERENT collaboration using the stopped-pion neutrino flux from the Spallation Neutron Source~\cite{cevns_ar} with a cross section on  order $22\cdot10^{-40} $cm$^2$ for an incoming average flux of $<E_\nu>\approx 30$ MeV from pion decay at rest.

\begin{figure}[ht]
        \centering
        \includegraphics[width=\textwidth]{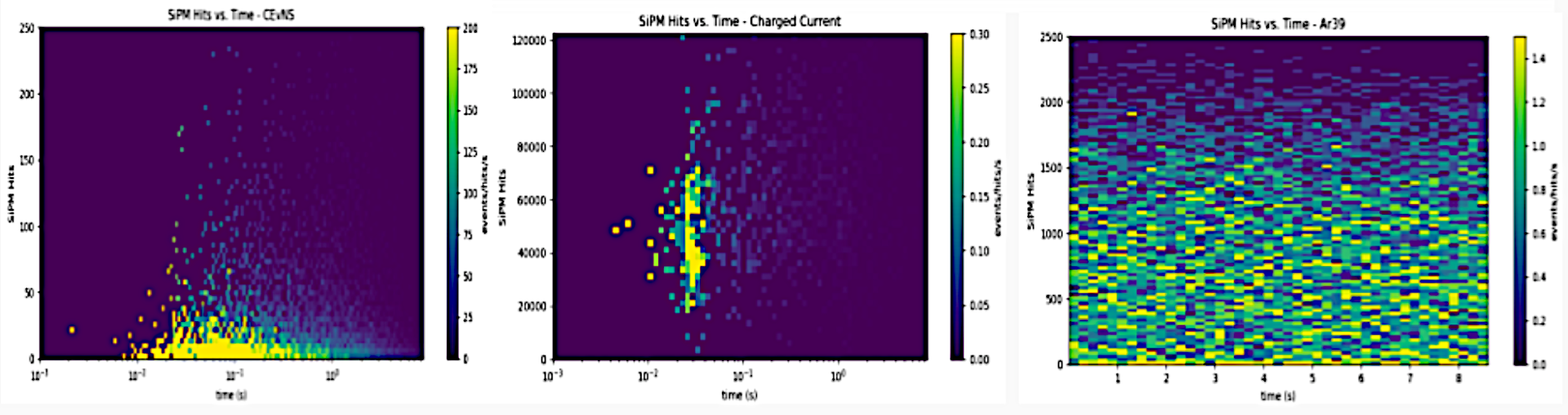}
        \caption{Distribution of SiPM hit counts as a function of time and multiplicity from a Geant4 simulation.  Left: CEvNS events for a 10-kpc core-collapse supernova. Center: $\nu_e$CC events from MARLEY. Right: $^{39}$Ar background~\cite{ortiz}. }
        \label{fig:sn_cevns_glow}
    \end{figure}

In a large LArTPC there will be a high rate of CEvNS in a core-collapse supernova burst--- approximately 30-100 times more events with respect to $\nu_e$CC (the dominant inelastic channel), depending on expected supernova spectrum.  However, each event is individually not likely to produce more than a few detected photons, and sub-50-keV thresholds need to be achieved to find these events, if they are to be found one by one. Even with depleted argon, the $^{39}$Ar rate down in this range in our proposed detector is by far overwhelming. However, an alternative is to identify a ``CEvNS glow,"~\cite{cevnsglow} in which the excess rate of detected photons can be tracked statistically above the $^{39}$Ar rate as a function of time in a characteristic explosion. Figure~\ref{fig:sn_cevns_glow} shows simulated supernova-induced activity from a burst at 10 kpc over 10 seconds in an inner fiducial 3 kT. One sees a  broadly-distributed-in-time, higher-energy charged-current (CC) component, the burst of low-hit-multiplicity CEvNS events at about 0.01 sec, and then the absolutely flat $^{39}$Ar activity. In ongoing work, we propose to subtract the reconstructed CC events and fit to the CEvNS bump above the background.   We note that in principle, an excess of collected ionization from CEvNS events is observable as a ``CEvNS buzz" in coincidence with the CEvNS photon glow.



\subsection{Solar Neutrino Physics}

In this section we present our enhanced sensitivity to solar neutrino searches including lower energies and enhanced oscillation sensitivity to $\Delta m^2_{21}$. This allows exploration of reactor-solar tensions and Non-Standard Interactions. It also allows a precision CNO solar neutrino measurement and a measurement of the $^3$He+p solar flux. A first study of DUNE as the next-generation solar neutrino experiment is in reference~\cite{PhysRevLett.123.131803}.


\subsubsection{Low Threshold Gains and Elastic scattering} \label{sec:low-threshold-es}

\begin{figure}[ht]
        \centering
        \includegraphics[width=0.7\textwidth]{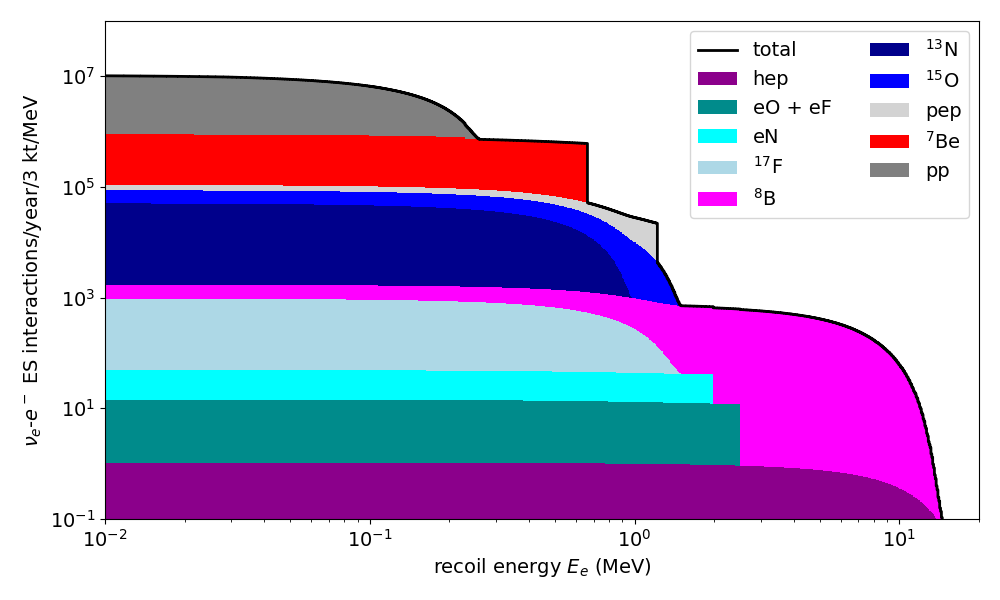}
        \caption{Differential solar-neutrino ES recoil spectrum in argon for a $3$-kt$\cdot$year exposure with contributions from different solar fluxes.}
        \label{fig:solar-es-recoils}
\end{figure}

One possible application for the lowered threshold would be detecting
solar neutrinos via elastic scattering on electrons (ES) in argon.
Figure~\ref{fig:solar-es-recoils} is the differential rate for this process
for a $3$-kt$\cdot$year exposure.
This plot shows that reducing
the threshold to $1$ MeV makes $pep$ and CNO neutrinos observable.
A threshold of $0.5$ MeV could add $^{7}$Be neutrinos,
and $0.1$ MeV could allow for detection of pp neutrinos.

\begin{figure}[ht]
        \centering
        \includegraphics[width=0.7\textwidth]{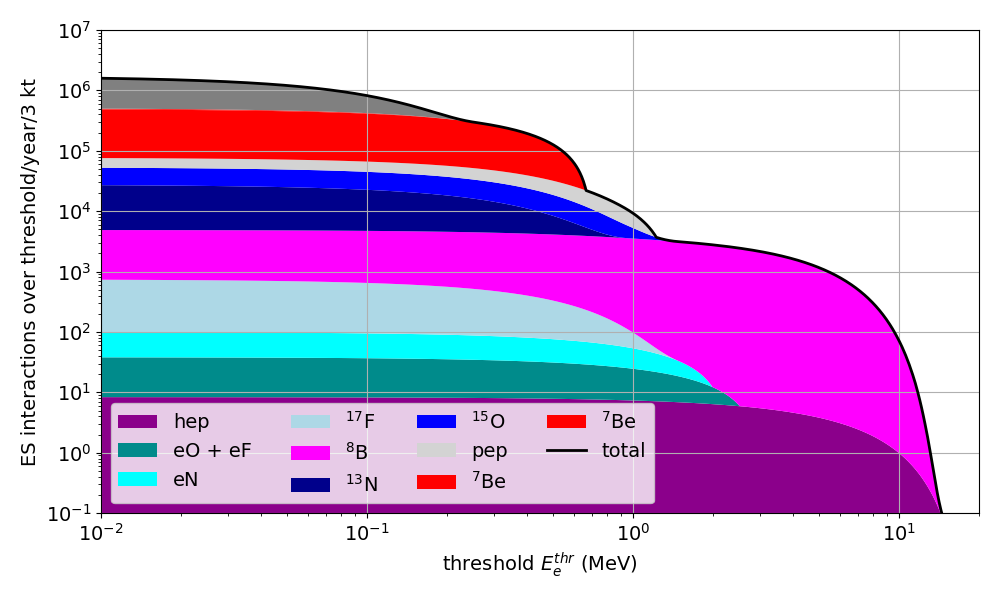}
        \caption{Number of events over threshold for solar-neutrino ES interactions in argon for a $3$-kt$\cdot$year exposure with contributions from different solar fluxes. }
        \label{fig:solar-es-over-threshold}
\end{figure}

Figure~\ref{fig:solar-es-over-threshold} shows the number of expected ES interactions over threshold
for a $3$-kt$\cdot$year exposure.
The total number of available neutrinos is important for possible
studies discussed in Section~\ref{sec:nsi}. This amounts to $9{,}200$ neutrinos
for a $1$-MeV threshold, $130{,}000$ neutrinos for a $0.5$-MeV threshold,
and $820{,}000$ for a $0.1$-MeV threshold.

In addition to the previously discussed methods for reducing backgrounds,
we are investigating directionality with ES for all solar neutrinos
and Cherenkov radiation for more energetic $^8$B neutrinos as ways
to enhance neutrino signal over backgrounds.


\subsubsection{Solar Neutrino Oscillations} 

Current measurements of the neutrino mixing parameter, $\Delta m^2_{21}$, using solar neutrinos from SNO and Super-Kamiokande are currently discrepant at 1.4~$\sigma$ with measurements from KamLAND using neutrinos from nuclear reactors~\cite{Esteban:2018ppq}.  Differing results from these two methods would suggest new physics possibly involved with exotic matter effects as the neutrino passes through the Sun and Earth.  Further data from DUNE will further investigate this discrepancy with high statistical significance.  The sensitivity comes from the ``day-night" effect, a partial regeneration of the $\nu_e$ solar flux due to matter effects in Earth which depends on $\Delta m^2_{21}$, neutrino energy, and nadir angle.  This is an advantageous strategy allowing for constraints of uncertainties using daytime data.

Solar neutrinos are much lower energy than typically observed in DUNE making reconstruction of these events difficult.  Also, at these low energies, radiological backgrounds, principally neutron capture with a contribution from $^{40}\text{Ar}(\alpha,\gamma)$, dominate analysis backgrounds.  A low-background DUNE-like module with enhanced light collection will help with both of these challenges.  

\begin{figure}[ht]
        \centering
        \includegraphics[width=0.75\textwidth]{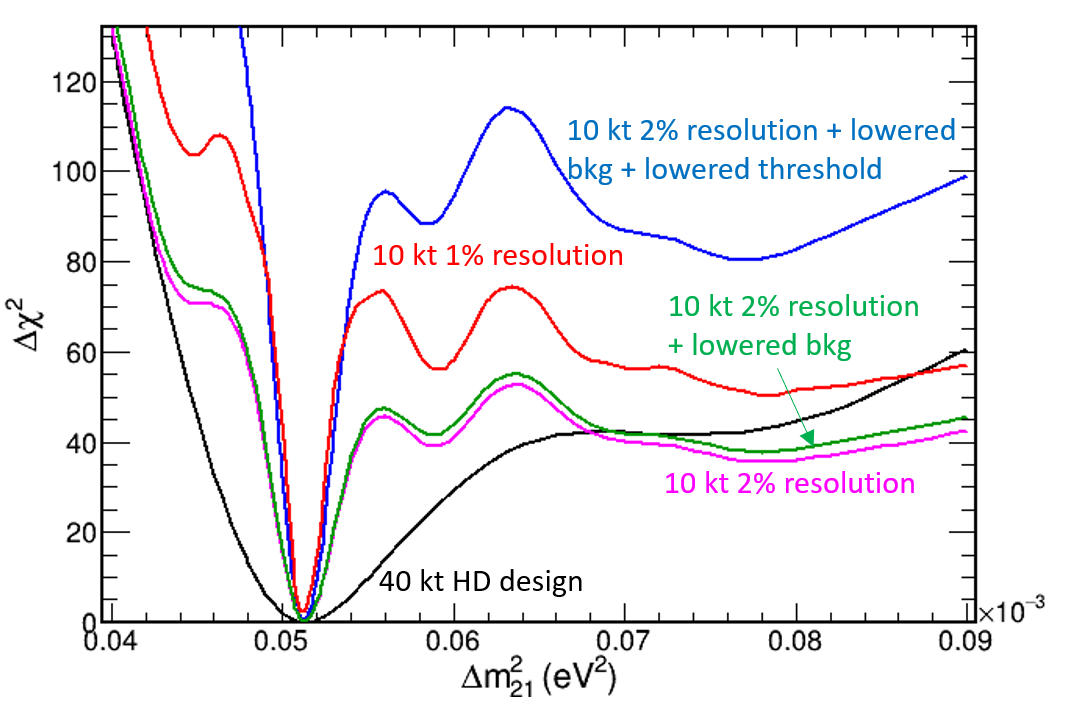}
        \caption{Sensitivity of a low-background module to the neutrino mixing parameter $\Delta m^2_{21}$ assuming a true value of the solar best fit, 5.13$\cdot 10^{-5}$~eV$^2$~\cite{Esteban:2018ppq}.  Colored contours show sensitivity after 100~kt-yrs for various detector configurations compared to 400~kt-yrs of DUNE data with horizontal drift design, shown in black. }
        \label{fig:sens_dmsq21}
\end{figure}

Improved energy resolution from increased photodetector coverage would significantly improve DUNE-like sensitivity to $\Delta m^2_{21}$.  This would both improve reconstruction of the dominant neutron background around the 6.1 MeV total visible energy of the neutron capture on argon-40, and better measure the dependence of the $\nu_e$ flux regeneration on neutrino energy.  A single 10-kt, low-background module could discern between SNO/SK and KamLAND best fits at over 6~$\sigma$.  Lower neutron background levels would also improve the signal-to-background ratio for the measurement, which could also lower the energy threshold for detecting and analyzing solar neutrino events.  An estimate of sensitivity to $\Delta m^2_{21}$ with 100~kt-yrs of exposure with a low-background module is compared to DUNE's sensitivity with 400~kt-yrs of data from nominal, horizontal drift modules in Fig.~\ref{fig:sens_dmsq21} with reasonable estimates of detector performance. We use a larger 10-kt fiducial volume for the reduced background/threshold curves in that figure, increased from other studies in this paper.

\subsubsection{Non-Standard Neutrino Interactions} \label{sec:nsi}

Non-standard neutrino interactions (NSI) could modify neutrino oscillations
in the Sun and result in a different number of neutrinos observed
compared to the one predicted by the Standard Model~\cite{Reichenbacher_2021}
(see also studies with different parameter definitions in~\cite{Friedland:2004pp} and~\cite{Friedland:2011za}).

\begin{figure}[ht]
        \centering
        \includegraphics[width=0.75\textwidth]{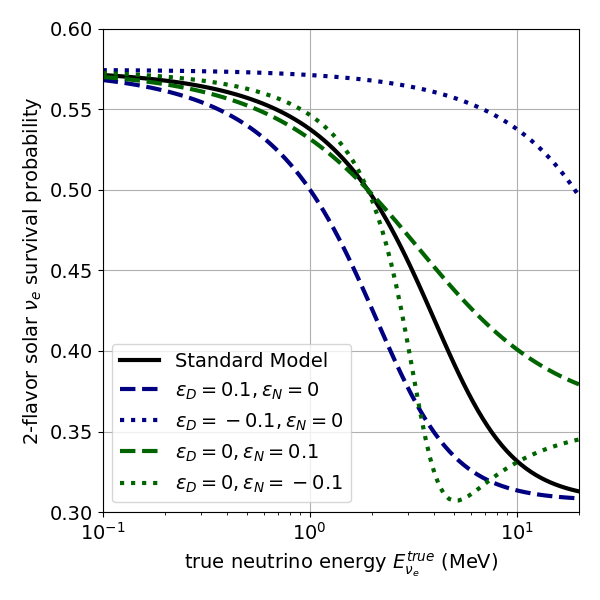}
        \caption{2-flavor electron-neutrino survival probability for solar neutrinos for the Standard Model and several different NSI couplings. }
        \label{fig:nsi-surv-prob}
\end{figure}

The NSI Hamiltonian (for neutral currents only) relevant for solar-neutrino oscillations can
be written in the following form:
\begin{equation}
H^{NSI}_\nu = \sqrt{2} G_F \left( n_u + n_d \right)
              \begin{pmatrix}
                  -\epsilon_D & \epsilon_N \\
                  \epsilon^{*}_N & \epsilon_D \\
              \end{pmatrix},
\end{equation}
where $G_F$ is the Fermi constant,
$n_u$ and $n_d$ are the up- and down-quark densities, respectively,
and $\epsilon_D$ and $\epsilon_N$ are the diagonal and off-diagonal
NSI couplings.
These couplings affect $\nu_e$ solar survival probability (see Figure~\ref{fig:nsi-surv-prob}).
The diagonal coupling can mimic different vacuum $\Delta$m$^2$ values,
resulting in its incorrect measurement when NSI are not taken into account.

Detection of ES in the proposed module (see Section~\ref{sec:low-threshold-es}) allows for a great
opportunity to have an almost NSI-independent anchor point in the oscillation probability near 0.1 MeV in addition to investigating NSI in
solar-neutrino oscillations at several MeV energies where changes in oscillation probability could be large but not much
experimental data exists, yet.

\begin{figure}[ht]
        \centering
        \includegraphics[width=0.65\textwidth]{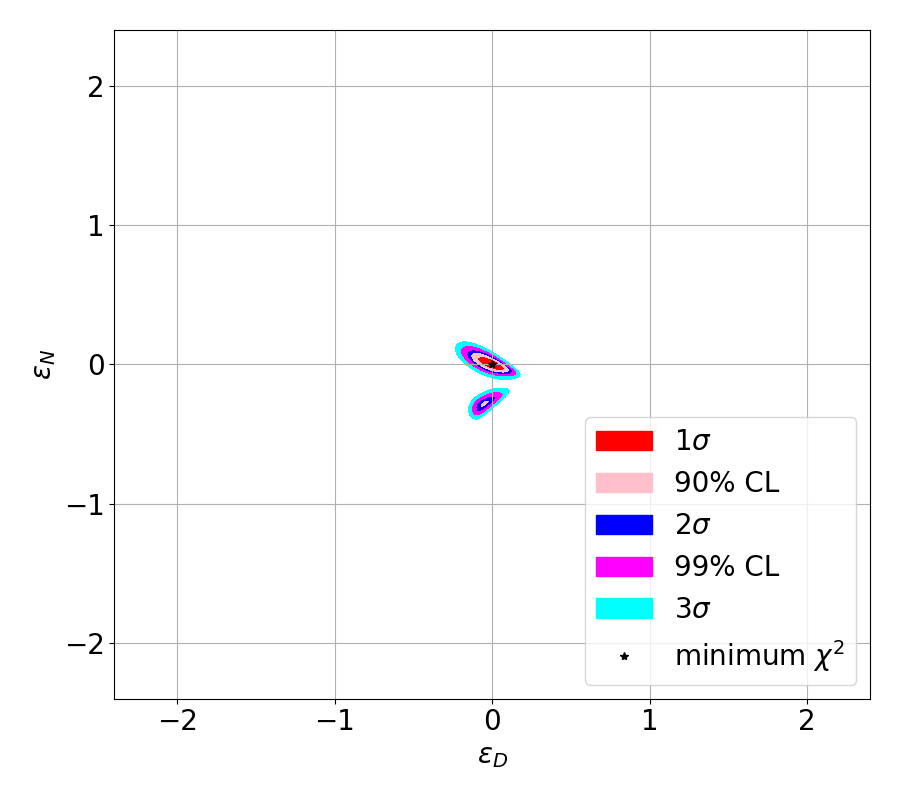}
        \caption{Possible NSI constraints for $3$ kt$\cdot$years obtained with the proposed module. }
        \label{fig:nsi-constraints}
\end{figure}

\begin{figure}[ht]
        \centering
        \includegraphics[width=0.65\textwidth]{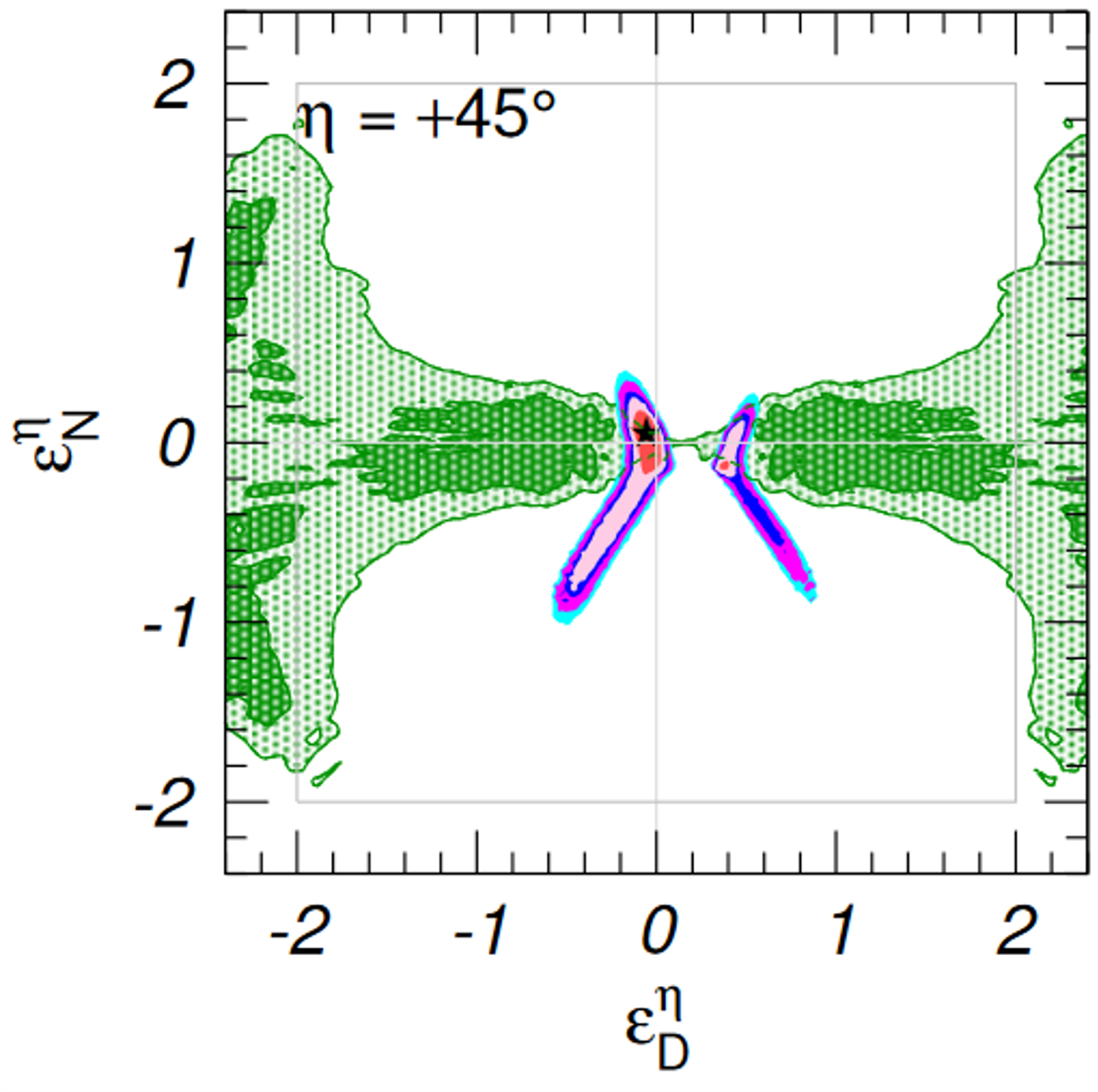}
        \caption{NSI constraints for similar NSI assumptions from~\cite{Esteban:2018ppq}
        (same contours as Figure~\ref{fig:nsi-constraints}). }
        \label{fig:nsi-current-constraints}
\end{figure}

An example of possible constraints on the NSI couplings is shown in Figure~\ref{fig:nsi-constraints}.
This plot was obtained with the assumption of no backgrounds or systematic errors,
an energy threshold of $1$ MeV, and an exposure of $3$ kt$\cdot$years. For comparison, recently obtained constraints using current neutrino data are available in~\cite{Esteban:2018ppq} and shown in Figure~\ref{fig:nsi-current-constraints} for a similar NSI model.

\subsubsection{Precision Measurement of CNO flux}

The CNO flux has been measured~\cite{cno} recently by the Borexino collaboration to 3.5$\sigma$ above 0, though it yields indistinct information about the high or low metallicity solution. Per ~\cite{cno}, "the CNO neutrino flux scales with the metal abundance in the solar core, itself a tracer of
the initial chemical composition of the Sun at the time of its formation. The metal abundance in
the core is thought to be decoupled from the surface by a radiative zone, where no mixing occurs.
CNO neutrinos are the only probe of that initial condition."

Here we want to investigate if an energy window exists where a precise measurement of the CNO flux can be performed in our low background module. We use our by-now standard simulation tools to count true deposited energy for a variety of backgrounds and solar neutrino sources in a 3 kT fiducial volume. We impose a 2\% energy smearing, as is reasonable by arguments in section~\ref{enressection}.  Fluxes come from~\cite{Vinyoles_2017} with a 0.3 survival probability applied, as appropriate to this low energy range. We use the assumption that the $^{42}$Ar content will be at a rate that is reduced by a further factor 100,000 compared to atmospheric argon. We emphasize again per section~\ref{sec:UAr} this is likely very conservative. The result is  we see in a window about 0.2 MeV wide just above the $pep$ cutoff that the CNO signal sticks up above other solar neutrino sources. This is merely an illustration;  a real analysis will likely do a templated fit above the $^7Be$ peak. Neutrons from the cold cryoskin stainless steel are forced to be low, as usual; radon is taken as controlled. The $^{210}$Bi background which Borexino took exquisite care to control and measure~\cite{cno} is yet to be thoroughly investigated here. Nevertheless, there would appear to be a window where the high and low metallicity solutions are  statistically separable. By comparison, CNO sensitivity in the two-phase LArTPC program, which studies are further along than the current work, can be seen in~\cite{Franco_2016}.

\paragraph{Triggering for CNO neutrinos}
Initial studies into measuring the CNO flux with TPC triggering are underway, taking into account the full complement of radiological backgrounds predicted in the DUNE detector.
In these early studies, TPC triggering requires sufficiently large, coincident clusters on the collection plane and at least one induction plane. We expect future work will in fact lead to a much higher-efficiency, light-based trigger being implemented for most work discussed in this paper. The dominant background in the 1.4 - 2.0 MeV energy window is radon in this early TPC triggering study, not in fact solar neutrinos -- despite the rate reduction by a factor of $10^{3}$ predicted in the low background module.
In order to perform the true CNO detection there is clearly much work to be done in  offline processing to accurately characterize and implement rejection by alpha detection, Bi-Po coincidence, disproportionate activity near the cathode from drifting cation decay products, and emanation properties of the various detector materials.

\begin{figure}[ht]
        \centering
        \includegraphics{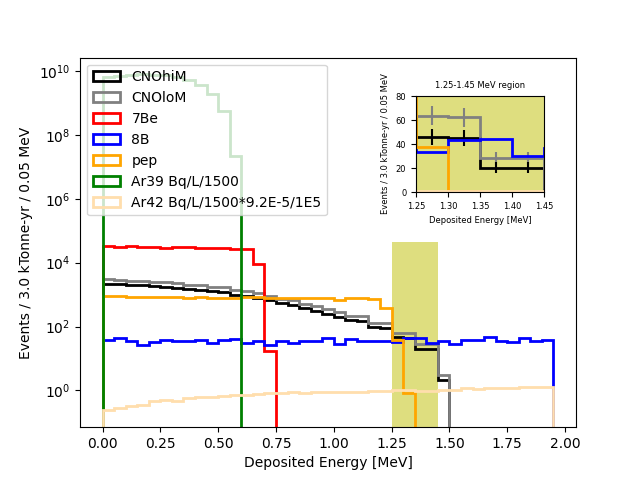}
        \caption{CNO solar neutrinos with backgrounds taking radon to be solved and negligible and neutrons constrained. This is 2\% smeared true deposited energy in our inner 3 kT fiducial volume in a year. Note that a likely, low $^{42}$Ar level is shown that reveals that a CNO fit is possible above the $^7Be$ peak and explicitly by a simple counting experiment in roughly the  region shown in yellow. This is a lower $^{42}$Ar level than shown in Fig 1, but still realistic. The inset  zooms in on the 1.25-1.45 above the $pep$ region to show that the signal statistics are large enough to favor either high- or low-metallicity after a year.}
        \label{fig:solarnu}
    \end{figure}

\subsubsection{Precision measurement of hep flux}

The hep solar neutrino process ($^3\mathrm{He} + p \rightarrow ^4\mathrm{He} + e^{+}+\nu_{e}$) produces the highest energy neutrinos, though they have the lowest flux and have not yet been observed. This low background module will be able to measure tens of these neutrinos per year, with a significant reduction in background due to radon and neutron interactions within the liquid argon.

\subsection{Neutrinoless Double Beta Decay}

A discovery of neutrinoless double beta decay (0$\nu2\beta$) is the most straightforward way to prove the Majorana nature of neutrinos. It would be  parsimonious to be able to search for 0$\nu2\beta$ in the same detector we are suggesting to use for the other measurements mentioned in this paper. $^{136}$Xe is one isotope in which much work is being performed worldwide to try to make this discovery~\cite{nexo,next}.

Loading large LArTPCs with few-percent level Xe and measuring neutrinoless double beta decay has been suggested in~\cite{snowmass_DUNE_beta} among other places. In this subsection we show that naively a discovery is likely possible with a signal $^{136}$Xe half-life of 5E28 years, and is quite apparent at 1E28 years, provided energy resolution requirements can be met. We imagine, say, a five year search campaign for 0$\nu2\beta$ at the end of the prosecution of the baseline physics program of this module, since any dark matter search requiring PSD would necessarily be compromised by xenon loading.

\begin{figure}[htpb]
    \centering
    \begin{subfigure}{}
        \centering
        \includegraphics[height=4in]{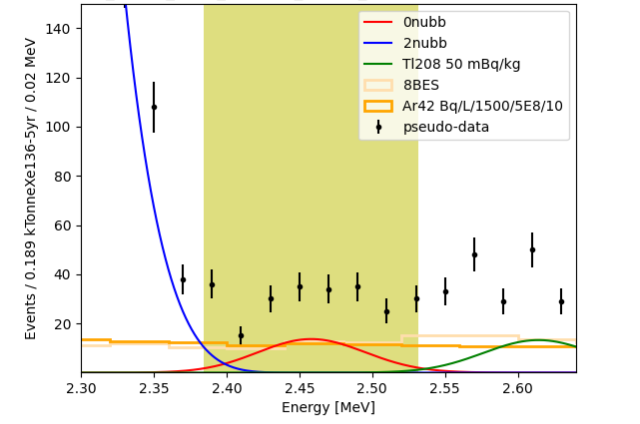}
    \end{subfigure}
    \begin{subfigure}
        \centering
        \includegraphics[height=4in]{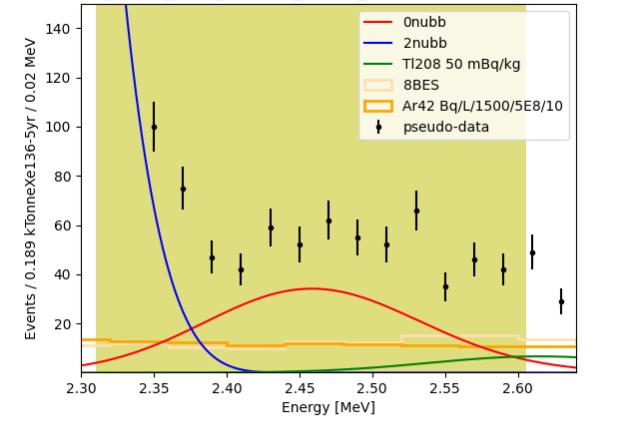}
    \end{subfigure}{}
    \caption{An optimistic background/resolution scenario for a $^{136}$Xe $0\nu\beta\beta$ half-life of 5E28 years. Detector energy resolution is 1.5\% on top. Backgrounds are as discussed in the text. On the bottom is the same with a reduced resolution of 3\% and for a half-life of 1E28 years.}
    \label{fig:0nubb-aggr}
\end{figure}

We start with a 0$\nu2\beta$ signal calculated using a Gaussian 1.5 (3) \% energy resolution  at 2.435 MeV with $1/\sqrt{E}$ dependence for top (bottom) plots in Figure~\ref{fig:0nubb-aggr}. Energy resolution ambitions are  consistent~\cite{lariat} with what we may expect in a LArTPC from charge energy collection  in combination  with the photon readout system, as well as discussion in section~\ref{enressection}. We  similarly smear the $2\nu$ double-beta true spectrum by these resolutions. And for the $^{208}$Tl background emanating from the G10 of the charge readout planes we use only the expected rate from the simulation, creating the actual spectrum by likewise smearing the 2.6 MeV gamma by 1.5 (3)\% in top (bottom) plots. For the solar and $^{42}$Ar backgrounds previously discussed, on the other hand, we use the resolution from the simulation that uses only the poorer light-only collection  from the SiPM hits. These are flat-ish backgrounds which extend to low energy where signal will likely be obscured in the noise of the charge readout, hence the need to rely on only the SiPMs. For both plots we use a 3\% concentration of $^{136}$Xe in our inner volume, using a 2.0 kTonne fiducial volume, and propose to run for five years. We plot a signal corresponding to a $(5) 1 \cdot 10^{28}$ year $0\nu\beta\beta$ half-life. 

Among the  assumptions made for our figures is that underground argon can give a $^{42}$Ar suppression of a factor of 10 beyond that in atmospheric Ar. See section~\ref{sec:UAr} which indicates this is likely easily achievable, up to processing concerns. We also take in Fig~\ref{fig:0nubb-aggr} that the irreducible solar $^8$B elastic scattering may be suppressed by a factor of two (conservatively down from three assumed in Reference~\cite{Brodsky:2018abk_cherenkov,ktonXe} ) from inspecting  Cernkov/Scintillation light ratios in single versus double electron events. We impose no efficiency hit due to this cut, whereas  Reference~\cite{Brodsky:2018abk_cherenkov} uses an efficiency of 0.75. More careful event reconstruction studies are needed to bear out the reasonableness of this cut. We assign the $^{208}$Tl in the G10 charge collection planes a Th concentration of 50 mBq/kg. Charged Current solar neutrino events, are in fact reducible to zero, due to the excited state gammas that are emitted. And similarly, neutrons shall be almost entirely removed using pulse-shape discrimination. We again take the radon issue to be solved for the sake of this study. A 2$\sigma$ band to either side of the $0\nu\beta\beta$ energy of 2.435 MeV is shown. 

We take  E$_{res}\sim 1.5$ \% in the top plot at the Q value, even though, as we have said, it is not immediately obvious we can achieve that (nor in fact are we currently confident we can reach the 2.5\% of the bottom plot) with our charge readout plus 80\% SiPM coverage, open as it is at the top and bottom. That study is beyond the scope of this paper. However, we see that there is sensitivity in this detector to $0\nu\beta\beta$  discovery at  lifetimes that stretch the reach of coming experiments, despite the large, irreducible solar $^8$B neutrinos which elastically scatter off the mostly argon target.


\subsection{Dark Matter}
\label{sec:darkmatter}

It is known that a large amount of dark matter exists within the Universe, that has so far only been observed by gravitation interactions. One popular candidate for the dark matter is the Weakly Interacting Massive Particle, or WIMP, that is the focus of several current and future experiments. The potential of using this low background module to search for WIMP dark matter was studied in Reference~\cite{DUNE-DM-PNNL}. This study assumed a dual phase TPC design, with a single fiducial volume at the center of the detector (rather than the split fiducial volumes described above). The criteria for achieving a sensitive WIMP dark matter search was set as requiring:
\begin{itemize}
    \item a 50-100 keV nuclear recoil threshold;
    \item $\mathcal{O}(10)$ background events;
    \item $\mathcal{O}(100)$ photons detected per event to allow pulse shape discrimination (PSD).
\end{itemize}

\begin{figure}[ht]
    \centering
    \includegraphics[width=8.5cm]{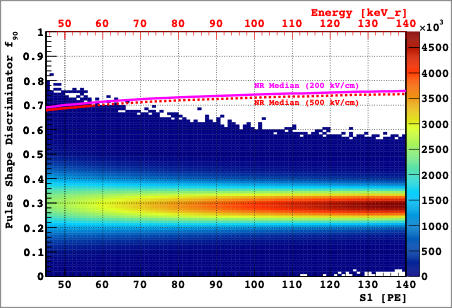}
    \caption{ Pulse shape discrimination F90 variable as a function of scintillation light S1. The figure shows the
39Ar electron recoil band simulated by the Monte Carlo simulation code described in the text. The median of
the nuclear recoil band at 200 V/cm and 500 V/cm (extrapolated above 57 keVnr ) is shown for comparison. Taken from~\cite{DUNE-DM-PNNL}}
    \label{fig:psd}
\end{figure}

Assuming that 1250 photons per 100 keV of prompt scintillation light is emitted (as measured by SCENE for 500 V/cm fields~\cite{PhysRevD.91.092007}), the studies in Section~\ref{optics} show that reaching 100 photons per event is realistic. A pseudo-Monte Carlo simulation of the Poisson distributed light output was performed to determine the width of a typical PSD variable f$_{90}$, defined as the ratio of light detector detected in the first 90 ns of an event to the light detected in the second 90 ns of an event, for the $^{39}$Ar decays. The means were take from measurements from the SCENE experiment~\cite{PhysRevD.91.092007}. This is shown in Figure~\ref{fig:psd}. PSD is expected to reach the 10$^{10}$ level.

All other electron/gamma backgrounds are expected to be subdominant to the $^{39}$Ar and will thus be removed by PSD. Neutron backgrounds were managed as described above. The main background will be from irreducible atmospheric neutrinos at the so-called neutrino floor. A full background table from the study is shown in Table~\ref{tab:bkgs}. 

\begin{figure}[ht]
        \centering
        \includegraphics[height=9.cm]{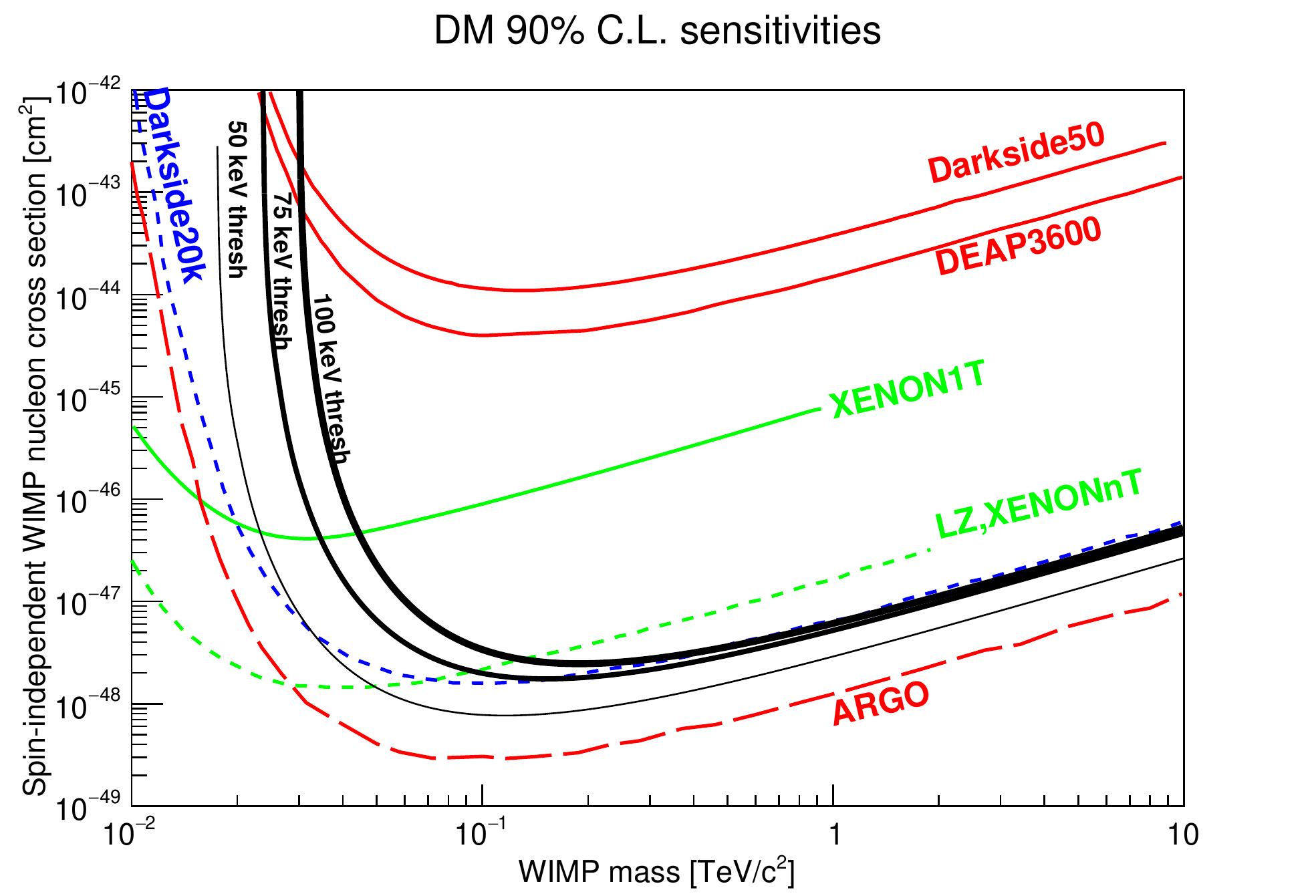}
        \caption{Shown are the achieved 90\% C.L. sensitivities in solid colored lines for various experiments for WIMP Dark Matter searches. Dashed lines for proposed argon and xenon experiments are overlaid. The DarkSide-20k exposure is 20 tons over a planned 10 years for a total of 200 ton-yrs, while our proposal for a DUNE-like module is 3 years of a 1 kt mass for a 3 kt-yr exposure. The ARGO exposure is a planned 300 tons over ten years for a total of 3 kt-yr. Two possibilities for a DUNE-like module 4 for two thresholds and expected backgrounds, are shown in black lines. A third, aspirational, DUNE module 4 sensitivity for a 50 keVnr threshold is also drawn as a thinner line. Taken from~\cite{DUNE-DM-PNNL}}
        \label{fig:sens}
    \end{figure}

The background rates are used to set a 90\% C.L. sensitivity to WIMP dark matter (see Figure~\ref{fig:sens}). This shows that a 3 year search with this detector will have comparable sensitivity to planned next generation detectors, which have expected run times of 10 years. This timescale allows a rapid cross-check of any signals discovered in these detectors, in particular for the liquid argon experiments such as DarkSide-20k~\cite{aalseth2018darkside} or ARGO~\cite{snowmass_gadmc}.  

\begin{table}[ht]
\centering
\begin{tabular}{ l c c c c c}
Background  & Amelioration strategy & & \multicolumn{3}{c}{Counts/3 kt-yr} \\
 & & & 100 keV$_r$ & 75 keV$_r$ & 50 keV$_r$\\
\hline
  neutrons from  & external 40 cm water & & 0.1 & 1.6 & 13  \\
  external rock & self-shielding, multi-site rej. & & & & \\[8pt]
 neutrons from & self-shielding, & & 1.02 & 14.2 & 2 \\
  cold cryoskin steel & acrylic, multi-site rej. & &  & & \\[8pt]
  $^{40}$K gammas & self-shielding, PSD & bPSD: & \multicolumn{3}{c}{$<4.3$} \\
  from detector top & & aPSD: & 0 & 0 & 0 \\[8pt]
$^{208}$Tl gammas & self-shielding, PSD & bPSD: & \multicolumn{3}{c}{$<30$} \\
from detector top & & aPSD: & 0 & 0 & 0\\[8pt]
$^{208}$Tl gammas & PSD & bPSD: & 8.1$\times10^4$ & 8.5$\times10^4$ & 8.9$\times10^4$ \\
from acrylic & & aPSD: & 0 & 0 & 0 \\[8pt]
$^{214}$Pb & PSD & bPSD: & \multicolumn{3}{c}{$< 1.9$ $\times10^8$} \\
 from radon & & aPSD: & 0 & 0 & 0 \\[8pt]
$^{40}$Ar $(\alpha,\text{n})$ & coincident tagging & & 0 & 0 & 0 \\
from radon& & & & & \\[8pt]
$^{39}$Ar betas & UAr, PSD & bPSD: & $1.6$$\times10^{10}$ & $1.7$$\times10^{10}$ & $1.8$$\times10^{10}$\\
in argon & & aPSD: & 0 & 1 & 1 \\[8pt]
atmospheric & none & & 10 & 13 & 17 \\
neutrinos & & & & &\\
\hline
Total & & & 11 & 30 & 33\\
\hline
\end{tabular}
    \caption{Backgrounds considered and estimated count rates in the inner 1 kt fiducial volume over 3 years for different energy thresholds. These estimates presume 40 cm water tanks between the rock wall and outer cryostat, the use of underground argon with an $^{39}$Ar rate of 7.3 × 10$^{−4}$ Bq/kgAr, multi-site rejection with 20~mm resolution, and a 50\% nuclear recoil acceptance. Electron recoil backgrounds are shown before (bPSD) and after (aPSD) application of PSD. Note, the numbers listed for a 50 keVr threshold require a change in the detector configuration. Taken from~\cite{DUNE-DM-PNNL}}
    \label{tab:bkgs}
\end{table}

\subsubsection{Seasonal Variation of Rate for WIMP Dark Matter}
\label{sec:AnnualWIMPmodulation}

\begin{figure}[ht]
        \centering
        \includegraphics[width=\textwidth]{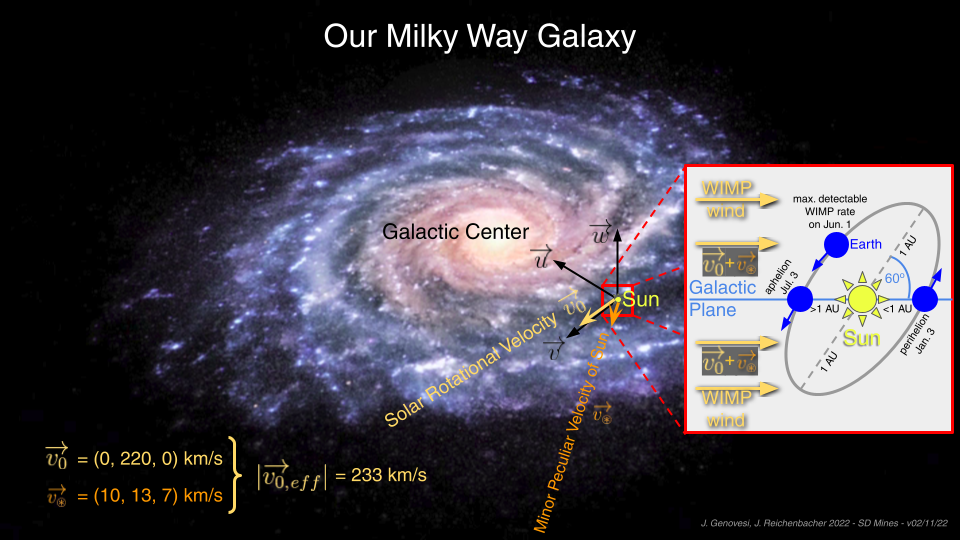}
        \caption{Galactic WIMP wind as it relates to Earth’s orbital plane employing an illustrative rendition \cite{SeasonalVariationModelRei21} of our Milky Way galaxy. Our solar system’s velocity in reference to our galaxy has contributions from both a rotational aspect with a tangential velocity of $220\,$km/s and from a minor solar peculiar aspect with velocity components $(U,V,W) = (10, 13, 7)\,$km/s \cite{SavageFreeseGondolo2006}. In our model the combined relative velocity of our solar system is then $233\,$km/s at which the Sun moves through a gas-like halo of WIMP dark matter assumed to have a Maxwell-Boltzmann-like velocity distribution. This induces what we experience as a "WIMP wind".  This WIMP wind is at an angle compared to Earth rotation around the Sun as pictured in the zoomed in diagram, with the effective WIMP velocity distribution shifted up when flowing maximally into the wind and shifted down when flowing maximally with the wind. Due to this aspect, an analysis of the annual modulation of detectable WIMP rate on Earth could provide a powerful tool for identifying the WIMP nature of dark matter.}
        \label{fig:AnnualModulationIllustration}
    \end{figure}

The prominent model for dark matter (DM) is the so-called Standard Halo Model (SHM) \cite{McCabe_2010} featuring WIMP DM. The abundance of DM derived from freeze out after the Big Bang roughly corresponds to the observed abundance, a coincidence that is often referred to as the "WIMP Miracle". The SHM describes a basic isometric spherical distribution of WIMP DM around our galaxy and has been used due to it being a good trade-off between realism and simplicity. The relative velocity of our solar system of $233\,$km/s \cite{SeasonalVariationModelRei21} at which the Sun moves through the gas-like halo of WIMP DM induces as a ”WIMP wind”. Figure \ref{fig:AnnualModulationIllustration} illustrates the Sun's rotation in our galaxy together with Earth's solar orbit into and out of the WIMP wind. We modeled the Sun's rotational and peculiar velocities into one combined effective velocity of $233\,$km/s in the galactic plane and accounted for the $60^{\circ}$ inclined plane of Earth's orbit. We were then able to accurately describe the annual modulation of detectable WIMP rate $R$ on Earth by one simple periodic sinusoidal function with one amplitude parameter $A$ and a maximal rate on June 1 of each year\cite{SeasonalVariationModelRei21} \cite{SavageFreeseGondolo2006}: 
\begin{equation}
\label{annualModulationFiteqn}
    R (\,[d^{-1}]\,) = A\,[d^{-1}] \,\times\, \cos\left( \frac{2\pi}{T[d]} \times (\,t[d] - t_{June1} [d]\,)\right) + R_{avg}\,[d^{-1}]
\end{equation}
Hereby the constant term $R_{avg}$ is the average annual rate. Earth's period $T$ is $365.2422\,$days and the phase corresponding to the maximal rate $R_{max}$ observable on June 1 of each year is $2\pi \cdot t_{June1}/T = 2\pi \cdot 0.415$. 

Galactic WIMPs in the halo are assumed to have a Maxwell-Boltzmann-like velocity distribution. The effective WIMP velocity distribution is shifted up when Earth is flowing maximally into the WIMP wind and shifted down when Earth is flowing maximally with the WIMP wind. Due to this aspect, an analysis on the annual modulation of the detection rate $R$ could provide a powerful tool for identifying WIMP DM. Due to the unrivaled large mass of 3 ktons and a potentially very long DUNE operation of one decade (or even several), this concept can offer a unique detection of the seasonal variation of the detectable WIMP rate $R$ at a sufficient statistical significance for providing a smoking gun signature for the WIMP nature of DM. This would be particularly of interest in case upcoming generation-2 DM experiments like LZ \cite{LUX-ZEPLIN:2018poe} and/or XENONnT \cite{XENON:2020kmp} have evidence for WIMPs near their sensitivity. It would be nearly impossible for the planned generation-3 DM experiments \cite{Aalbers:2022dzr} \cite{Bottino:2022zky} to make such a smoking gun detection proving the WIMP nature of DM. 

\begin{figure}{ht}
        \centering
        \includegraphics[width=5.5in]{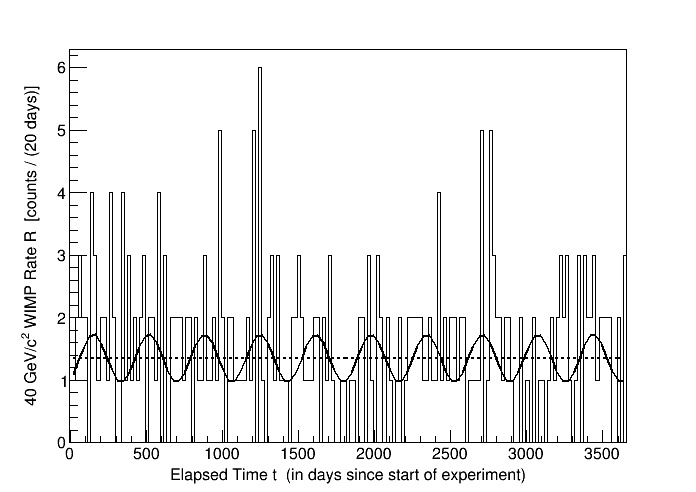}
        \includegraphics[width=5.5in]{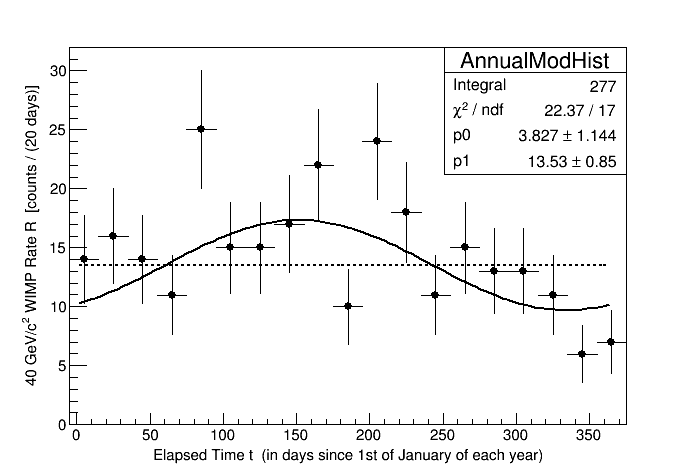}
        \caption{Exemple NEST\cite{Szydagis:2021hfh} simulation of the annual modulation for $40\,$GeV/$c^2$ WIMP dark matter with a cross section of $4\times10^{-48}\,$cm$^2$ for a 10 year measurement time with $3\,$kton LAr at $50\,$keV threshold. On top a Likelihood-fit result for a 10 year period and on bottom the same data combined into a single annual period starting on January 1 of each year fitted with a $\chi^2$-method. The ideal case without background is assumed in this study.}
        \label{fig:AnnualModulationResult}
    \end{figure}

The differential rate of interactions in an arbitrary detector for the SHM is described in Equation \ref{totalDifferentialRate}:

\begin{equation}
\label{totalDifferentialRate}
\frac{dR}{dE_R} = \sigma_N^{SI} \frac{A^2 m_A N_T \rho_{\chi}}{2 m_{\chi} \mu^2_N} F^{2}(E_R) \int\limits_{{\nu}_{min}(E_R)}^{\infty} \frac{d^{3}\overrightarrow{\nu}}{\nu} f_{\oplus}(\overrightarrow{\nu},\overrightarrow{\nu_{obs}})
\end{equation}
where $\sigma_N^{SI}$ is the spin-independent-nucleon cross-section for WIMPs, $A$ is the atomic number of argon, $m_A$ is the mass of argon, $N_T$ is the number of target nuclei, $\rho_{\chi}$ is the local dark matter density ($0.3 \frac{GeV}{cm^{3}}$), $m_{\chi}$ is the mass of a WIMP, $\mu_N$ is the reduced WIMP-nucleus mass, $F(E_R)$ is the the nuclear form-factor, $\nu_{min}$ is minimum WIMP velocity to produce a recoil of energy $E_R$, $\nu$ is WIMP velocity, and $\nu_{obs}$ is the observer velocity with respect to the galaxy. When Earth is moving into or with the WIMP wind, it affects the differential WIMP rate, which in turn would affect our $\overrightarrow{\nu_{obs}}$. For ease, we can define:

\begin{equation}
\int\limits_{{\nu}_{min}(E_R)}^{\infty} \frac{d^{3}\overrightarrow{\nu}}{\nu} f_{\oplus}(\overrightarrow{\nu},\overrightarrow{\nu_{obs}}) = {\zeta}(E_R),
\end{equation}
where
\begin{equation}
\label{zetaToBe}
f(\overrightarrow{\nu}) = \frac{1}{N}\left(e^{\frac{-{\nu}^{2}}{\nu_0^{2}}} - e^{\frac{-\nu_{esc}^{2}}{\nu_0^{2}}}\right),
\end{equation}
and 
\begin{equation}
f_{\oplus}(\overrightarrow{\nu},\overrightarrow{\nu_{obs}}) = f(\overrightarrow{\nu} +\overrightarrow{\nu_{obs}})
\end{equation}

with N as a normalization factor, $\nu_{0}$ is the expected WIMP velocity, and $\nu_{esc}$ is the escape velocity of the galaxy. Using Equation \ref{zetaToBe}, we can solve for individual cases of WIMPs in certain speed brackets. 





As mentioned earlier, Earth's orbit can play a crucial role for differential WIMP rate predictions in the SHM. As the solar system travels through our galaxy, observers on Earth would observe WIMPs dominantly from a certain direction in a windshield-to-rain like effect. This is due to the distribution of WIMPs being treated as a gas with a Maxwell-Boltzmann-like velocity distribution with the stars in our galaxy moving through the dark matter due to their orbits around the galactic nucleus. In addition to the standard rotational contribution from the solar system, a minor peculiar velocity exists of our solar system traveling through the galaxy. This WIMP wind would be at an angle of $60^{\circ}$ towards Earth's orbital plane. This means that during June 1, the Earth will experience a maximum effective flux of WIMPs while on December 1 the Earth will have experienced the lowest effective WIMP rate, as one can see again in Figure \ref{fig:AnnualModulationIllustration}. 

An example fit using Equation \ref{annualModulationFiteqn} of an annual modulation signal simulated in NEST \cite{Szydagis:2021hfh} without background is shown in Figure~\ref{fig:AnnualModulationResult} on top for a 10 year period, and on bottom the same data combined into a single annual period starting on January 1 of each year. This is for a $40\,$GeV/$c^2$ WIMP with a cross section of $4\times10^{-48}\,$cm$^2$, which is very close to the limit of sensitivity of the upcoming generation-2 xenon dark matter experiments LZ \cite{LUX-ZEPLIN:2018poe} and XENONnT \cite{XENON:2020kmp}. We further assumed a $3\,$kton $\times$ $10\,$year exposure of our proposed low background LAr module with a $50\,$keV threshold resulting in $277$ events. Bottom of Figure~\ref{fig:AnnualModulationResult} shows the $\chi^2$-fit result of the amplitude parameter $A$ ($=\,$p0) of Equation \ref{annualModulationFiteqn}. It confirms the possible measurement of the seasonal variation of the WIMP rate at a sufficient statistical significance for providing a smoking gun signature for the WIMP nature of DM. This will be uniquely possible with this detector, due to the unrivaled large mass of $3\,$ktons vs. only $300\,$ tons of argon for ARGO \cite{Bottino:2022zky} and $100\,$ tons for a Gen-3 xenon experiment \cite{Aalbers:2022dzr}. Moreover, the annual modulation effect in xenon is significantly smaller due to the relatively lower energies of nuclear recoils in xenon compared to argon. Last but not least, the logistics of a decade long operation with this detector can utilize strong synergies with the DUNE long-baseline physics, including the cavern availability and occupancy.

\subsection{Additional Topics}



This module, with its unprecedented combination of low background and size, also can explore several other topics. We describe several examples in this section.

\subsubsection{Atmospheric neutrinos}
The detector will measure approximately 10 CE$\nu$NS events due to atmospheric neutrinos. These events have not yet been observed and this would allow a cross-check of background rates from the upcoming generation of dark matter experiments.

\subsubsection{Strangelets}
Recently, the paper "Can strangelets be detected in a large LAr neutrino detector?" \cite{Parvu:2021lrt}, predicted that a LArTPC detector is able to detect and discriminate light strangelets with (Z, A) between (2,14) and (7,70) for energies up to 10 GeV in the presence of radioactive background found at the surface. When operated underground the detection limits are expected to be extended due to lower background levels and, combined with the increased dimensions of the detector module, will improve the event rates with a factor of 40. In the case of strangelets the main uncertainties are due to the estimations of their survival probability deep underground. The presence of $^{39}$Ar masks both ionization and scintillation signals from strangelets and induces false signals in the collected charge from ionization. The use of underground argon in this module allows a cleaner detection signal.

\subsubsection{Charged micro-black holes}
Hawking \cite{Hawking:1971ei} suggested that unidentified tracks in the photographs taken in old bubble chamber detectors could be explained as signals of gravitationally "collapsed objects" ($\mu$BH). The small black holes are expected to be unstable due to Hawking radiation, but the evaporation is not well-understood at masses of the order of the Planck scale. Certain inflationary models naturally assume the formation of a large number of small black holes \cite{Chen:2004ft} and the generalized uncertainty principle may indeed prevent total evaporation of small black holes by dynamics and not by symmetry, just like the hydrogen atom is prevented from collapse by the standard uncertainty principle \cite{Adler:2001vs}. Given the profound nature of the issues addressed, some disagreement and controversy exist.

In principle the direct detection of charged micro black holes with masses around and upward of the Planck scale (10$^{-5}$ g), ensuring a classical gravitational treatment of these objects, is possible in huge LAr detectors. It has been shown that the signals (ionization and scintillation) produced in LAr enable the discrimination between micro black holes (with masses between 10$^{-5}$ - 10$^{-4}$ grams, and velocities in the range 250 - 1000 km/s) and other particles \cite{Lazanu:2020qod}. It is expected that the trajectories of these micro black holes will appear as crossing the whole active medium, in any direction, producing uniform ionization and scintillation on the whole path.

In the direct detection of the charged micro-black holes, unlike in traditional WIMP detection, there will exist both ionization and scintillation signals from direct interactions and from recoiling nuclei. The capability to perform pulse shape discrimination in this detector will allow these tracks to be identified. Natural radioactivity is the main source of background in this case and the reduced number of free electrons (and photons) from beta decays of $^{39}$Ar will allow a significant improvement of the capability of the detector to correctly identify the micro-black hole signals.

\subsubsection{Other topics}
This detector would have sensitivity to a small number of CE$\nu$NS events within the neutrino beam. It may have applications to geologic tomography. The improved energy resolution for low energy could have applications in searches for other exotics and beyond the standard model physics phenomena. Though the optimal search region for a diffuse supernova neutrino background is above the energy of the solar neutrinos~\cite{M_ller_2018}, and thus the radioactive backgrounds, the improved energy resolution of this detector will again likely improve the search sensitivity. 

\section{Conclusion}

We have presented a design for low background kton-scale detector, based on the vertical drift design planned for DUNE, which could be a candidate for a third or fourth DUNE Module of Opportunity. This can be achieved by:
\begin{itemize}
    \item additional shielding;
    \item stringent radioactive background control;
    \item and enhanced light detection.
\end{itemize}
We have also described some of the physics searches that could be enhanced with this detector including improved sensitivity to supernova neutrinos, solar neutrinos, neutrinoless double beta decay and WIMP dark matter.

\bibliographystyle{unsrt} 
\bibliography{lbgd}




\end{document}